\RequirePackage{etex}
\renewcommand{\raggedright}{}
\documentclass[man, 12pt, longtable, floatsintext]{apa7}

\usepackage[utf8]{inputenc}
\DeclareUnicodeCharacter{2212}{\textminus} 

\usepackage[american]{babel}

\usepackage{amsmath, amssymb}
\usepackage{url}
\usepackage{bbold}
\usepackage{graphicx}
\usepackage{csquotes}
\usepackage{algorithm}
\usepackage{algpseudocode}

\usepackage[normalem]{ulem}

\usepackage{autonum}
\RequirePackage{etoolbox}

\definecolor{BlueViolet}{HTML}{473992}
\hypersetup{
plainpages=false,%
bookmarksopen=true,%
bookmarksnumbered=false,%
bookmarksdepth=5,
breaklinks=true,
colorlinks=true,
citecolor=BlueViolet,
urlcolor=blue,
}
\makeatletter
\newcommand{\blx@noerroretextools}{}
\makeatother
\usepackage[style=apa, 
backend=biber, 
uniquename=false,
url=false,
doi=false]{biblatex}
\usepackage{bm}
\usepackage{bbm}
\usepackage{mathrsfs} 
\usepackage{mathtools}
\usepackage{amsmath}
\usepackage{amssymb}
\usepackage{amsfonts}
\allowdisplaybreaks[4] 

\usepackage{siunitx}
\usepackage{mathrsfs}

\usepackage{mleftright}
\newcommand{\maru}[1]{\mleft(#1\mright)}

\newcommand{\kagi}[1]{\mleft[#1\mright]} %bracket

\newcommand{\nami}[1]{\mleft\{#1\mright\}} %curly bracket

\newcommand{\abs}[1]{\mleft\vert#1\mright\vert}

\newcommand{\btA}{{\mathbf{A}}}

\newcommand{\btD}{{\mathbf{D}}}

\newcommand{\btQ}{{\mathbf{Q}}}

\newcommand{\btY}{{\mathbf{Y}}}

\newcommand{\bta}{{\mathbf{a}}}

\newcommand{\btd}{{\mathbf{d}}}

\newcommand{\calN}{{\mathcal{N}}}

\newcommand{\bbR}{{\mathbb{R}}}

\newcommand{\bTheta}{{\bm\Theta}}
\newcommand{\btheta}{{\bm\theta}}

\newcommand{\bSigma}{{\bm{\Sigma}}}

\title{Sparse Bayesian joint modal estimation for exploratory item factor analysis}
\shorttitle{Sparse Bayesian joint modal estimation for exploratory IFA}

\authorsnames[1,2,1]{Keiichiro Hijikata, Motonori Oka, Kensuke Okada}
\authorsaffiliations{{The University of Tokyo},{London School of Economics and Political Science}}
\authornote{Correspondence concerning this article should be addressed to Keiichiro Hijikata, The University of Tokyo, 7-3-1 Hongo, Bunkyo-ku, Tokyo, 113-0033 Japan. Email: k.hijikata.1120@outlook.jp}

\leftheader{Hijikata et al.}

\journal{JEBS}

\abstract{
This study presents a scalable Bayesian estimation algorithm for sparse estimation in exploratory item factor analysis based on a classical Bayesian estimation method, namely Bayesian joint modal estimation (BJME). BJME jointly estimates the model parameters and factor scores by maximizing the complete-data joint posterior density. The proposed algorithm achieves scalability through an alternating optimization scheme that iteratively updates the model parameters and factor scores. Simulation studies show that the proposed algorithm has high computational efficiency and high accuracy in variable selection over latent factors and parameter recovery. Moreover, we provide a real data analysis using a large-scale dataset from a psychological assessment of the Big Five personality traits. The results indicate that the proposed algorithm extracts an interpretable factor loading structure in a computationally efficient manner.}


\date{\today}

\begin{document}
\maketitle

\section{Introduction}

Item factor analysis \parencite[IFA;][]{wirth2007item} has been widely used to analyze categorical data consisting of many observed variables, primarily in psychological and educational measurement. Exploratory IFA reveals the relationships between observed variables (e.g., items) and latent variables (e.g., factors), which are captured through the magnitude of factor loadings.
%\sout{observed variables, i.e., items, and a smaller number of latent variables, i.e., factors, which are represented as factor loadings}\cred{observed variables (e.g., items) and latent variables (e.g., factors), which are captured through the magnitude of factor loadings.} 
When revealing such relationships, it is desirable to obtain factor loadings in a simple structure; that is, ‘‘each factor pertained as much as possible to one nonoverlapping subset of the observed variables'' \parencite[][p.8]{mulaik2009foundations}
%\sout{for estimated factor loadings to be sparse}\cred{to obtain factor loadings in a simple structure; that is, ‘‘each factor pertained as much as possible to one nonoverlapping subset of the observed variables \parencite[][p.8]{mulaik2009foundations}} 
for a better interpretation of the relationships 
%\cred{between observed and latent variables}. 
between observed and latent variables. Researchers have developed two approaches for finding 
%\sout{the sparse} \cred{a simple} 
a simple structure of factor loadings: rotation and regularization. In the rotation approach, factor loadings are estimated with some regular identification constraints and then rotated under some rotation criterion, e.g., varimax \parencite{kaiser1958varimax} and quartimin \parencite{carroll1953analytical}. On the other hand, the regularization approach typically employs an $L_1$-norm type penalty function and shrinks the estimates whose absolute values are small toward zero \parencite[e.g.,][]{sun2016latent,cho2022regularized}. Consequently, the loadings of the irrelevant factors are reduced to zero, and a sparse latent structure can be extracted.
%, leading to a simple structure of factor loadings.

% test design的にsimple structureになるよおうにitemを作る、そうしたitemから生まれたデータをスパース推定すればsimple structureをrecoverできるはずだ。

Comparing these two approaches, the issue with the rotation technique is that it cannot necessarily produce sufficiently sparse estimates \parencite{hirose2015sparse}. This is because it does not make some irrelevant loadings zero, but only attempts to make them close to zero within the matrix rotation framework.
%but finds factor loadings with a simple structure where the absolute values of irrelevant factor loadings are small but not necessarily zero.
% Furthermore, \sout{that can lead to a subjective interpretation as the rotated factor loadings need a cutoff based on a subjective thresholding value, such as $0.3$ and $0.4$.} \cred{as the rotated factor loadings need a cutoff to obtain a simple structure of factor loadings based on a subjectively-chosen thresholding value, such as $0.3$ and $0.4$ \parencite{cho2022regularized}, the resulting structure of factor loadings can depend on such thresholding values.}
% as the rotated factor loadings need a cutoff to obtain a simple structure of factor loadings based on a subjectively chosen thresholding value, such as $0.3$ and $0.4$ \parencite{cho2022regularized}, the resulting structure of factor loadings can depend on such thresholding values.
In contrast, 
%\sout{the regularization approach can overcome these issues since some of the elements of factor loadings by the regularized estimation become exactly zero.}\cred{the regularization approach can avoid such an issue since it can allow the factor loadings of irrelevant factors to shrink to exactly zero with the optimally-tuned value of penalty.} 
the regularization approach can alleviate this issue because it allows the factor loadings of irrelevant factors to shrink to exactly or approximately zero with the optimally tuned penalty weight. Therefore, this study adopted a regularization approach for exploratory IFA. %Furthermore, although there are several models in IFA, such as the multidimensional two-parameter logistic model, our focus 
In this study, we focused specifically on % is placed on 
the multidimensional graded response model \parencite[MGRM; e.g.,][]{muraki1995full,cai2010high}
as the model for exploratory IFA.
This choice was motivated by the prevalence of
%In this study, we chose the MGRM as the model for exploratory IFA because IFA often involves 
categorical item responses %rather than binary ones, 
such as those found in survey questionnaires and psychological assessments, %with a Likert scale. Thus, choosing the MGRM for exploratory IFA is aligned with the common data type 
in the applied settings. % of the relevant fields. 

% bayesian ifaのレビュー 
% Birnbaum (1968) 
% review 
We adopt the Bayesian inference framework to perform statistical inference in the sparse estimation for exploratory IFA. There are a number of studies focusing on the Bayesian IFA and its related models in the literature. For example, \textcite{fox2010bayesian} discussed various Bayesian item response models. \textcite{albert1992bayesian,beguin2001mcmc} proposed a Markov chain Monte Carlo (MCMC) algorithm for the Bayesian IFA. \textcite{muthen2012bayesian} developed an MCMC algorithm for Bayesian structural equation modeling. \textcite{arminger1998bayesian} proposed a Bayesian approach to estimate nonlinear latent variable models using an MCMC algorithm. 
Although MCMC methods are commonly used in Bayesian inference, as mentioned above, they are known to be computationally intensive under large-scale settings where the numbers of latent variables, respondents, and items are all large. In particular, modern psychological and educational measurements typically comprise large-scale item response data with numerous respondents and items owing to the increasing availability of computer-based assessments and online platforms for data collection. The emergence of this kind of large-scale data has led to a growing demand for computationally efficient alternatives to MCMC methods.

Researchers have recently resorted to variational Bayesian (VB) methods as a computationally efficient alternative to MCMC methods. Examples of VB applications in psychometrics can be found in both fully Bayesian estimation \parencite[e.g.,][]{natesan_bayesian_2016,wu2021modeling} and marginal maximum likelihood estimation \parencite[MMLE; e.g.,][]{cho2021gaussian,jeon_variational_2017}. In the case of MMLE, VB methods are applied to the posterior distribution of the latent variables to derive the lower bound of the log marginal likelihood. VB methods approximate an intractable posterior distribution with a family of probability distributions (e.g., exponential family) that is more tractable than the true posterior, which is called the variational distribution. They find the parameters of the variational posterior distribution that is close to the corresponding true posterior by minimizing the Kullback–Leibler divergence between these two distributions. 

Although VB methods have received attention within the psychometric community over the past decade owing to their methodological advances in the fields of machine learning and statistics, this study revisits a classical approach to Bayesian estimation called \emph{Bayesian joint modal estimation} \parencite[BJME;][]{swaminathan1982bayesian,swaminathan1985bayesian} and sheds light on its utility as a computationally efficient alternative to MCMC methods. 
BJME finds the estimates of the model parameters and factor scores that maximize the complete-data joint posterior density constructed based on the joint likelihood of the model parameters and factor scores. Thus, it can be viewed as a Bayesian counterpart of recently developed joint maximum likelihood estimation (JMLE) methods \parencite[e.g.,][]{chen2019joint,chen2020structured,gu2023joint,bergner2022multidimensional,lan2014sparse}.
In the IFA literature, \textcite{swaminathan1982bayesian, swaminathan1985bayesian} proposed BJME algorithms for the Rasch and unidimensional two-parameter logistic models, respectively.
VB methods familiar to the psychometric community, such as coordinate ascent variational inference, typically require model-specific derivations and tricks to obtain conditional conjugacy by using, for example, a local variational method \parencite{jaakkola_bayesian_2000}. However, BJME requires only the computation of gradients with respect to the parameters of interest. This means that its optimization can be easily addressed by a gradient method and thus handled in a computationally efficient manner, such as in JMLE. Furthermore, because of its simple nature in optimization using a gradient method, one can also easily incorporate inequality constraints or non-smooth penalties on the parameters, such as a monotonicity constraint or $L_1$-norm penalty, which are commonly employed in models used in IFA, by capitalizing on a proximal update. This flexibility is intriguing when it comes to performing regularized estimation in exploratory IFA.

Accordingly, we adopt the BJME method and propose its sparse estimation algorithm for exploratory IFA under the MGRM. In particular, the proposed sparse BJME method for the exploratory MGRM can be performed by utilizing a Laplace (double-exponential) prior over factor loading parameters. This is because the negative log of the Laplace prior is proportional to the $L_1$-norm penalty function \parencite{tibshirani1996regression}. With the optimal choice of the parameter of the Laplace prior, the loadings of irrelevant factors can be shrunk to zero. Similarly, we can assign the normal prior over the other parameters of interest, namely, factor scores and category intercept parameters in the MGRM, where the normal prior turns out to be the squared $L_2$-norm penalty function. Using the correspondence between the priors and penalty functions, the sparse estimation for the exploratory MGRM can be performed in a simple and computationally efficient manner through a gradient ascent method with a proximal update.
In addition, from this connection between the prior and penalty, the BJME framework can be understood not only as a Bayesian counterpart to JMLE methods but also directly as a penalized JMLE framework. Within this framework, maximizing the log complete-data joint posterior is equivalent to maximizing a joint log-likelihood subject to penalties.

The remainder of this paper is organized as follows. First, we briefly introduce the MGRM and its Bayesian formulation. Subsequently, the BJME algorithm is proposed in detail. Second, we conduct simulation studies to evaluate the estimation accuracy and computational efficiency of the proposed algorithm in comparison with the MCMC and MMLE methods. Third, we apply the proposed algorithm to a real large-scale dataset of psychological assessments to measure the Big Five factors. Finally, we summarize the findings of this study and discuss the limitations and future extensions of the proposed method.

\section{Model}\label{sec:2}
\subsection{Multidimensional graded response model}
This section introduces the MGRM. Note that the indices for respondents, items, and factors are $i\;(=1,\ldots, N)$, $j\;(=1,\ldots, J)$, and $k\;(=1,\ldots, K)$, respectively. The number of response categories for item $j$ is denoted by $C_j$. Let the observed data be $\btY = (Y_{ij})_{N \times J}$, where $Y_{ij}\in\{0,\dots,C_j-1\}$. We denote a vector of factor scores, factor loadings, and category intercepts by $\btheta_i = (\theta_{i1}, \ldots, \theta_{iK})^\top$, $\bta_j =(a_{j1}, \ldots, a_{jK})^\top$, and $\btd_j = (d_{j1}, \ldots, d_{j,C_j-1})^\top$, respectively. Here, $\top$ is used to denote the transpose. With these notations, we set $\bTheta = (\btheta_1, \ldots, \btheta_N)^\top$, $\btA = (\bta_1, \ldots, \bta_J)^\top$, and $\btD = (\btd_1, \ldots, \btd_J)^\top$. Furthermore, we assume $d_{j1} > \ldots > d_{j,C_j-1}$ for model identification.

Next, the MGRM considers the following response probabilities of $Y_{ij}$:
\begin{align}
P(Y_{ij}\geq0 \mid \boldsymbol{\theta}_i,\mathbf{a}_j,\mathbf{d}_j)&=1, \\ 
P(Y_{ij}\geq1 \mid\boldsymbol{\theta}_i,\mathbf{a}_j,\mathbf{d}_j)&=\text{logit}^{-1}(\boldsymbol{\theta}_i^{\top}\mathbf{a}_j+d_{j1}), \\
\vdots \nonumber \\
P(Y_{ij}\geq C_j-1\mid\boldsymbol{\theta}_i,\mathbf{a}_j,\mathbf{d}_j)&=\text{logit}^{-1} (\boldsymbol{\theta}_i^{\top}\mathbf{a}_j+d_{j,C_j-1}), \\
P(Y_{ij}\geq C_j\mid\boldsymbol{\theta}_i,\mathbf{a}_j,\mathbf{d}_j)&=0,
\end{align}
where $\text{logit}^{-1}(\cdot)$ is the inverse logit function: $\text{logit}^{-1}(z)=({1+\exp(-z)})^{-1}$ for $z\in\mathbb{R}$. Accordingly, the probability that the response $Y_{ij}$ takes a certain category can be modeled as
\begin{align}\label{eq:resprob}
P(Y_{ij}=c\mid\boldsymbol{\theta}_i,\mathbf{a}_j,\mathbf{d}_j)=P(Y_{ij}\geq c \mid\boldsymbol{\theta}_i,\mathbf{a}_j,\mathbf{d}_j )-P(Y_{ij}\geq c+1 \mid\boldsymbol{\theta}_i,\mathbf{a}_j,\mathbf{d}_j),
\end{align}
for $c=0,\dots,C_j-1$. 

\subsection{Fully Bayesian representation of the MGRM}

Here, we present a fully Bayesian representation of the MGRM. First, using Equation~\ref{eq:resprob} and assuming that respondents are independent and identically distributed and that items are locally independent given factor scores, the complete-data likelihood function of the MGRM is written as
\begin{align}
    P(\btY, \bTheta \mid \btA, \btD ;\mathbf{0}_K, \bSigma_\theta) 
    &= P(\btY \mid \bTheta, \btA, \btD)P(\bTheta \mid \mathbf{0}_K, \bSigma_\theta) \\
    &= \prod_{i=1}^N \kagi{ \prod_{j=1}^J\prod_{c=0}^{C_j-1}P(Y_{ij}=c \mid \btheta_i, \bta_j, \btd_j)^{\mathbb{1}{\nami{Y_{ij}=c}}} P(\btheta_i\mid \mathbf{0}_K, \bSigma_\theta)},
\end{align}
where the parameters after the semicolon ($;$) denote the ones that are prespecified and not estimated; $\mathbb{1}{\nami{\cdot}}$ is the indicator function. The prior distribution of $\btheta_i$ is the multivariate normal distribution given as
\begin{align}
    P(\btheta_i
    \mid \mathbf{0}_K, \bSigma_\theta) = \frac{\det\maru{\bSigma_{\theta}}^{-1/2}}{\maru{2\pi}^{K/2}}\exp\nami{-\frac{1}{2}\btheta_i^\top\bSigma_{\theta}^{-1}\btheta_i}, \label{equ: theta prior}
\end{align}
where we set the mean vector of $\btheta_i$ to the $K$-dimensional zero vector, $\mathbf{0}_K$; $\bSigma_\theta$ is the covariance matrix of factor scores. In this study, we do not estimate $\bSigma_\theta$ but prespecify the estimate obtained prior to parameter estimation. In the context of sparse inference for exploratory IFA, handling the factor covariance matrix $\bSigma_\theta$ can involve either joint estimation or prespecification; for instance, \textcite{sun2016latent} discussed both alternatives and opted for using a prespecified $\bSigma_\theta$ in their main analysis. Following a similar path, we adopt the prespecification strategy in this study. The joint distribution of the prior over $\btheta_i$ is then given as $P(\bTheta\mid \mathbf{0}_K, \bSigma_\theta) = \prod_{i=1}^N P(\btheta_i\mid \mathbf{0}_K, \bSigma_\theta)$. 

Second, the prior distribution of $a_{jk}$ is specified as the Laplace distribution:
\begin{align}
    P(a_{jk}\mid 0, \lambda^{-1}) = \frac{\lambda}{2}\exp\nami{-\lambda \abs{a_{jk}}}, \label{equ: a prior}
\end{align}
where we set its mean to 0; $\lambda^{-1}>0$ is a scale parameter of the Laplace distribution that controls the degree of sparsity for factor loadings. Although the value of $\lambda$ could differ for each $j$, we let $\lambda$ be the same across all $j$. This choice is made for the ease of tuning $\lambda$. Our simulation and empirical studies show that the proposed algorithm maintains good performance in sparse estimation even with the same value of $\lambda$ across all $j$. 
The joint distribution of the prior over $a_{jk}$ is then given as $P(\btA\mid 0, \lambda^{-1}) = \prod_{j=1}^JP(\bta_j\mid 0, \lambda^{-1}) = \prod_{j=1}^J \prod_{k=1}^K P(a_{jk}\mid 0, \lambda^{-1})$. 

Third, the prior distribution of $d_{jc}$ is the normal distribution given as
\begin{align}
    P(d_{jc}\mid 0, \sigma_d^2)= \frac{1}{\sqrt{2\pi \sigma_d^2}}\exp\nami{-\frac{1}{2\sigma_d^2}d_{jc}^2 }, \label{equ: d prior}
\end{align}
where we set its mean to 0 and $\sigma_d^2=100^2$ to assign a diffuse prior over $d_{jc}$. 
The joint distribution of the prior over $d_{jc}$ is then given as $P(\btD\mid 0, \sigma_d^2) = \prod_{j=1}^JP(\btd_j \mid 0, \sigma_d^2) = \prod_{j=1}^J \prod_{c=1}^{C_j-1} P(d_{jc}\mid 0, \sigma_d^2)$.

Given the complete-data likelihood function and priors, the complete-data joint posterior distribution is given as
\begin{align}\label{eq:cd_jointposterior}
P(\bTheta, \btA, \btD &\mid \btY; \nami{\mathbf{0}_K,\bSigma_\theta}, \nami{0, \lambda^{-1}},\nami{0, \sigma_d^2}) \\ &\propto P(\btY, \bTheta \mid \btA, \btD ;\mathbf{0}_K, \bSigma_\theta)P(\btA\mid 0, \lambda^{-1})P(\btD\mid 0, \sigma_d^2),
\end{align}
where curly brackets $\nami{\cdot}$ are used to denote a set of prespecified hyperparameters in the corresponding prior distributions.
The complete-data joint posterior in Equation \ref{eq:cd_jointposterior} is typically constructed by the data augmentation (DA) algorithm using some MCMC sampler \parencite[][Chapter 10.1]{brooks2011handbook}, and there, the target posterior is the $\nami{\btA, \btD}$-marginal of the complete-data joint posterior because 
it is characterized by the likelihood marginalized over the latent variables.
%the likelihood function is characterized by the marginal likelihood that involves marginalizing out latent variables.
The posterior distribution of latent variables (i.e., $\bTheta$) is then obtained as a byproduct of the DA algorithm. In contrast, BJME directly maximizes the posterior kernel of the complete-data joint posterior, which is constructed based on the joint likelihood of the model parameters and factor scores. For this reason, the BJME can be considered as the Bayesian counterpart of the JMLE that targets the joint likelihood of the model parameters and factor scores instead of the marginalized likelihood over latent variables for its optimization in parameter estimation. 

\section{Method} \label{sec:3}
\subsection{Algorithm}

The BJME method finds the estimates of the model parameters and factor scores that maximize the complete-data joint posterior density in Equation~\ref{eq:cd_jointposterior},
which is based on the joint likelihood of the model parameters and factor scores. In the following expositions, $P(\boldsymbol{\Theta},\mathbf{A},\mathbf{D}\mid\mathbf{Y})$ denotes the complete-data joint posterior based on the joint likelihood for notational simplicity.

The objective function in this optimization problem is the logarithm of the posterior kernel for the complete-data joint posterior density based on the joint likelihood, which is given as
\begin{align}
\ell(\bTheta, \btA, \btD\mid \btY, \lambda)\ &= \sum_{i=1}^N {\sum_{j=1}^J \sum_{c=0}^{C_j-1} \mathbbm{1}\nami{Y_{ij}=c}\log P(Y_{ij}=c \mid \btheta_i, \bta_j, \btd_j)} \\
&\quad + \sum_{i=1}^N \log P(\btheta_i \mid \mathbf{0}_K, \bSigma_\theta)  + \sum_{j=1}^J \log P(\bta_j \mid 0, \lambda^{-1})  + \sum_{j=1}^J \log P(\btd_j \mid 0, \sigma_d^2). \label{eq:objfun}
\end{align}
To jointly seek the estimates $(\hat{\bTheta},\hat{\btA},\hat{\btD})$ that maximize the objective function in Equation~\ref{eq:objfun}, we develop an alternating optimization algorithm \parencite[e.g.,][]{chen2019joint}, where each of $(\boldsymbol{\Theta},\mathbf{A},\mathbf{D})$ is alternately updated until convergence is reached. 

First, the log prior density for each parameter can be expressed from Equations~\ref{equ: theta prior}, \ref{equ: a prior}, and \ref{equ: d prior} as 
\begin{align}
\log P(\boldsymbol{\theta}_i\mid\mathbf{0}_K,\boldsymbol{\Sigma}_\theta) &= -\frac{K}{2}\log (2\pi) -\frac{1}{2}\log\det\maru{\bSigma_\theta}-\frac{1}{2}\boldsymbol{\theta}_i^{\top}\boldsymbol{\Sigma}_{\theta}^{-1}\boldsymbol{\theta}_i \label{equ:theta prior density}, \\
\log P(\mathbf{a}_j\mid0,\lambda^{-1}) &= K\log \left(\frac{\lambda}{2}\right)-\lambda\|\mathbf{a}_j\|_1 \label{equ:a prior density}, \\
\log P(\mathbf{d}_j\mid0,\sigma_d^2) &= -\frac{C_j - 1}{2}\log (2\pi\sigma_d^2)-\frac{1}{2\sigma_d^2}\|\mathbf{d}_j\|^2 \label{equ:d prior density}.
\end{align}
$\|\cdot\|_1$ and $\|\cdot\|$ represent the $L_1$- and $L_2$-norms, respectively. 

Second, the objective function in Equation \ref{eq:objfun} is smooth with respect to both $\boldsymbol{\theta}_i$ and $\mathbf{d}_j$. Therefore, they can be updated using the gradient ascent method. It is noted that since parameter $\mathbf{d}_j$ is constrained to be strictly ordered, we reparameterize $\mathbf{d}_j$ as $\boldsymbol{\delta}_j = (d_{j1},\log(d_{j1}-d_{j2}),\dots,\log(d_{j,C_j-2}-d_{j,C_j-1}))^\top$ 
during the optimization process. The objective function in Equation \ref{eq:objfun} is not smooth with regard to $\mathbf{a}_j$ due to the $L_1$-norm term in Equation~\ref{equ:a prior density}. Accordingly, we employ the proximal gradient method to update parameter $\mathbf{a}_j$.

Finally, the updating rules for the model parameters and factor scores are represented as
\begin{align}
\boldsymbol{\theta}_i^{[t+1]} &\gets \boldsymbol{\theta}_i^{[t]} + \gamma \frac{\partial\ell(\bTheta^{[t]}, \btA^{[t]}, \btD^{[t]} \mid \btY, \lambda)}{\partial \boldsymbol{\theta}_i^{[t]}}, \text{ for all $i$},\label{equ:theta update} \\
\mathbf{a}_j^{[t+1]} &\gets S_{\lambda\gamma}\left(\mathbf{a}_j^{[t]} + \gamma \frac{\partial\log P( \btY \mid \bTheta^{[t+1]}, \btA^{[t]}, \btD^{[t]})}{\partial \mathbf{a}_j^{[t]}}\right),\text{ for all $j$}, \label{equ:a update}\\
\boldsymbol{\delta}_j^{[t+1]} &\gets \boldsymbol{\delta}_j^{[t]} + \gamma \frac{\partial\ell(\bTheta^{[t+1]}, \btA^{[t]}, \btD^{[t]}\mid \btY, \lambda)}{\partial \boldsymbol{\delta}_j^{[t]}}, \text{ for all $j$}
\label{equ:d update},
\end{align}
where the superscript $t$ is the iteration number. Furthermore, $\gamma$ denotes the step size determined by line search in each step. $S_{\lambda\gamma}(\cdot)$ is the element-wise soft thresholding operator defined as, for $\mathbf{z} \in \bbR^K$,  
\begin{align}
    S_{\lambda\gamma}(\mathbf{z})=\begin{pmatrix}
    \textrm{sign}(z_1)\max(|z_1|-\lambda\gamma,0) \\
    \vdots \\
    \textrm{sign}(z_K)\max(|z_K|-\lambda\gamma,0)
    \end{pmatrix}.
\end{align}
The function $\log P( \btY \mid \bTheta^{[t+1]}, \btA^{[t]}, \btD^{[t]})$ in Equation~\ref{equ:a update} represents the log joint likelihood term in the objective function:
\begin{align}
\log P( \btY \mid \bTheta^{[t+1]}, \btA^{[t]}, \btD^{[t]}) = \sum_{i=1}^N {\sum_{j=1}^J \sum_{c=0}^{C_j-1} \mathbbm{1}\nami{Y_{ij}=c}\log P(Y_{ij}=c \mid \btheta_i^{[t+1]}, \bta_j^{[t]}, \btd_j^{[t]})}.
\end{align}
Regarding intercepts $\mathbf{d}_j$, we obtain $\mathbf{d}_j^{[t+1]}$ applying the inverse transformation to $\boldsymbol{\delta}_j^{[t+1]}$ as $d_{jc}^{[t+1]} = \delta_{j1}^{[t+1]} - \sum_{c^\prime=2}^{c} \exp(\delta_{jc^\prime}^{[t+1]})$, where the sum is taken to be zero if $c=1$.
The detailed derivations of the gradients are provided in Appendix.

For the convergence criterion, we stop running the algorithm if the absolute inter-iteration change in the objective function drops below 5. 
If $\mathbf{Y}$ contains missing entries, we replace them with zero and set the likelihood contribution from the missing entries to zero. 

\subsection{Parallel computing}
The proposed algorithm can be implemented in parallel because of the assumptions that respondents are independently distributed and items are locally independent given factor scores. This allows the updating rules to be defined independently for each $i$ and $j$. In our implementation, we split $\{1,\dots,N\}$ and $\{1,\dots,J\}$ into equal-sized subsets. Accordingly, the number of subsets is set to the number of cores used for parallel computing. Subsequently, we perform parameter updates for each subset on each core. Parallel computing is implemented using the \texttt{Distributed} package in the Julia language \parencite[][]{bezanson2017julia}. Our algorithm for obtaining $(\hat{\boldsymbol{\Theta}},\hat{\mathbf{A}},\hat{\mathbf{D}})$ can be summarized in Algorithm~\ref{pseudocode}.
\begin{algorithm}[htp]
\caption{Sparse BJME algorithm for the MGRM}
\label{pseudocode}
\algdef{SE}[PARFOR]{ParFor}{EndParFor}[1]{\textbf{for #1 do in parallel}}{\textbf{end for}}
\begin{algorithmic}[1]
\Statex \hspace*{-\algorithmicindent} \textrm{Input:} $\mathbf{Y},K,\boldsymbol{\Sigma}_\theta,\lambda,\sigma_d^2$
\Statex \hspace*{-\algorithmicindent} \textrm{Output:} $\hat{\boldsymbol{\Theta}},\hat{\mathbf{A}},\hat{\mathbf{D}}$
\State {Initialize $(\boldsymbol{\Theta}^{[1]},\mathbf{A}^{[1]},\mathbf{D}^{[1]})$. For example, for all $i, j, k, c$:
    \begin{itemize}
        \item[] \hspace{-0.6em} Draw $\boldsymbol{\theta}_i^{[1]}$ from a multivariate standard normal distribution, $\mathcal{N}(\mathbf{0}_K, \mathbf{I}_K)$
        \item[] \hspace{-0.6em} Draw $a_{jk}^{[1]}$ from a uniform distribution, $\textrm{Uniform}(0, 2)$
        \item[] \hspace{-0.6em} Draw $d_{jc}^{[1]}$ from $[-1.5, 1.5]$ subject to the ordering constraint
    \end{itemize}
}
\State Set $t \gets 1$
\Repeat
    \ParFor {$i\gets1,\dots,N$}
        \State {Obtain $\boldsymbol{\theta}_i^{[t+1]}$ according to  Equation~\ref{equ:theta update}}
    \EndParFor
    \ParFor {$j\gets1,\dots,J$}
        \State{Reparameterize $\mathbf{d}_j^{[t]}$ as $\boldsymbol{\delta}_j^{[t]}$}
        \State{Obtain $(\mathbf{a}_j^{[t+1]},\boldsymbol{\delta}_j^{[t+1]})$ according to  Equations~\ref{equ:a update} and \ref{equ:d update}}
        \State{Obtain $\mathbf{d}_j^{[t+1]}$ from $\boldsymbol{\delta}_j^{[t+1]}$ through the inverse transformation}
    \EndParFor
    \State Set $t \gets t+1$
\Until{the absolute inter-iteration change in the objective function is less than 5}

\State {$\hat{\boldsymbol{\Theta}},\hat{\mathbf{A}},\hat{\mathbf{D}} \gets \boldsymbol{\Theta}^{[t]},\mathbf{A}^{[t]},\mathbf{D}^{[t]}$}
\end{algorithmic}
\end{algorithm}

\subsection{Selection of penalty weight}
Here, we describe a cross-validation (CV) procedure to select the value of $\lambda$. The procedure consists of data-splitting \parencite[e.g.,][]{fithian2017optimal} and two-stage selection \parencite[e.g.,][]{lee2010sparse}. We first randomly divide the dataset into training and test sets. The training set is used to perform a two-stage selection of $\lambda$: a coarse search over a prespecified set of candidate values and a refined search within a narrower interval. The final value of $\lambda$ is then used to estimate the model parameters on the test set. We state this procedure in more detail below.

First, we randomly partition the rows of $\mathbf{Y}$ into a training set $\mathbf{Y}_1$ and a test set $\mathbf{Y}_2$. Given the training set $\mathbf{Y}_1=(Y_{i_1j})_{N_1\times J}$, where $N_1$ indicates the number of respondents in the training set, we perform a two-stage selection of $\lambda$ using a missing-value-based CV method \parencite[e.g.,][]{chen2019joint}. Let $\boldsymbol{\Omega}_1=(\omega_{i_1j})_{N_1\times J}$ be the indicator matrix of observed responses in the training set $\mathbf{Y}_1$; that is, $\omega_{i_1j}=1$ if $Y_{i_1j}$ is observed and $\omega_{i_1j}=0$ otherwise. We randomly split $\boldsymbol{\Omega}_1$ into $M$ mutually exclusive sets that are of approximately equal size: $\boldsymbol{\Omega}_1^{(m)}=(\omega_{i_1j}^{(m)})_{N_1\times J},m=1,\dots,M$. Thus, they satisfy $\sum_{m=1}^M \boldsymbol{\Omega}_1^{(m)}=\boldsymbol{\Omega}_1$. Here, we use $M$ to indicate the number of folds, and we set $M=5$ in this study.

We perform two-stage selection based on $\mathbf{Y}_1$ and $\boldsymbol{\Omega}_1^{(m)}=(\omega_{i_1j}^{(m)})_{N_1\times J},m=1,\dots,M$. In the first stage, we evaluate the CV prediction error for a prespecified set of coarsely spaced candidate values, $\lambda^{(1)}=\{0.01,0.1,1,10,100\}$. We fit the proposed algorithm to the training set $\mathbf{Y}_1$ on $\boldsymbol{\Omega}_1^{(-m)}$ for $m=1, \dots, M$, where $\boldsymbol{\Omega}^{(-m)}_1=\sum_{m^\prime\neq m}\boldsymbol{\Omega}^{(m^\prime)}_1$ indicates the data excluding the set $m$. Then, the CV prediction error is calculated as
\begin{equation} \label{equ:cv fn}
\textrm{CV}(\lambda)=\sum_{m=1}^M \sum_{i_1,j:\:\omega_{i_1j}^{(m)}=1}\sum_{c=0}^{C_j-1} -\mathbbm{1}\nami{Y_{i_1 j} = c }\log P(Y_{i_1j}=c\mid\hat{\boldsymbol{\theta}}_{i_1}^{(m)},\hat{\mathbf{a}}_j^{(m)},\hat{\mathbf{d}}_j^{(m)}),
\end{equation}
where $\boldsymbol{\Omega}^{(m)}_1$ is used for out-of-sample evaluation and $(\hat{\boldsymbol{\theta}}_{i_1}^{(m)},\hat{\mathbf{a}}_j^{(m)},\hat{\mathbf{d}}_j^{(m)})$ are the estimates obtained by fitting the proposed algorithm on $\boldsymbol{\Omega}^{(-m)}_1$. We select the value $\hat{\lambda}^{(1)}$ that minimizes the prediction error among the candidate values of $\lambda^{(1)}$.

In the second stage, we refine the selection of $\lambda$ by searching within a narrower range around $\hat{\lambda}^{(1)}$. We define a new set of candidate values $\lambda^{(2)}$, where the values are more densely spaced within the interval $[\lambda_\textrm{min}^{(2)}, \lambda_\textrm{max}^{(2)}]$. The boundaries of this interval are determined based on $\hat{\lambda}^{(1)}$ and a prespecified positive integer $R$, which controls the width of the search range, as $\lambda_\textrm{min}^{(2)} = \max(0, \hat{\lambda}^{(1)} / R)$ and $\lambda_\textrm{max}^{(2)} = R\hat{\lambda}^{(1)}$. Then, we generate $R$ equally spaced values within this range and set $\lambda^{(2)}$ to these values. In this study, we set $R=5$. Note that the integer $R$ is unrelated to the number of CV folds $M$; although both are set to five in this study, they serve different purposes. We repeat the $M$-fold CV process with the set $\lambda^{(2)}$ and select the final penalty weight $\hat{\lambda}$ that minimizes the prediction error among $\lambda^{(2)}$. Finally, we fit the proposed algorithm to the test set $\mathbf{Y}_2$ under $\hat{\lambda}$. This CV procedure is summarized in Algorithm~\ref{alg:cv}.
\begin{algorithm}[htp]
\caption{Cross-validation procedure for penalty weight selection}
\label{alg:cv}
\begin{algorithmic}[1]
\Statex \hspace*{-\algorithmicindent} \textrm{Input:} $\mathbf{Y}, M, R, \lambda^{(1)}$
\Statex \hspace*{-\algorithmicindent} \textrm{Output:} $\hat{\lambda}$

\State Randomly split $\mathbf{Y}$ into training set $\mathbf{Y}_1$ and test set $\mathbf{Y}_2$

\State Randomly partition the indicator matrix $\boldsymbol{\Omega}_1$ into $M$ sets: $\boldsymbol{\Omega}_1^{(1)},\dots,\boldsymbol{\Omega}_1^{(M)}$

\State \texttt{// Stage 1: coarse search}
\For{each $\lambda \in \lambda^{(1)}$}
    \For{$m \leftarrow 1,\dots,M$}
        \State Fit the proposed algorithm using $\boldsymbol{\Omega}_1^{(-m)}$ as the observed-entry indicator
        % \State Compute the CV prediction error on $\boldsymbol{\Omega}_1^{(m)}$
    \EndFor
    \State Compute the CV prediction error $\textrm{CV}(\lambda)$ according to Equation~\ref{equ:cv fn}
\EndFor

\State Select $\hat{\lambda}^{(1)}=\arg\min_{\lambda\in\lambda^{(1)}} \textrm{CV}(\lambda)$

\State \texttt{// Stage 2: refined search}
\State Set $\lambda_{\min}^{(2)}=\max(0,\hat{\lambda}^{(1)}/R)$ and $\lambda_{\max}^{(2)}=R\hat{\lambda}^{(1)}$

\State Generate $R$ equally spaced values in $[\lambda_{\min}^{(2)},\lambda_{\max}^{(2)}]$ and denote them by $\lambda^{(2)}$

\For{each $\lambda \in \lambda^{(2)}$}
    \For{$m \leftarrow 1,\dots,M$}
        \State Fit the proposed algorithm using $\boldsymbol{\Omega}_1^{(-m)}$ as the observed-entry indicator
        % \State Compute the prediction error on $\boldsymbol{\Omega}_1^{(m)}$
    \EndFor
    \State Compute the CV prediction error $\textrm{CV}(\lambda)$ according to Equation~\ref{equ:cv fn}
\EndFor

\State Select $\hat{\lambda}=\arg\min_{\lambda\in\lambda^{(2)}} \textrm{CV}(\lambda)$

% \State Fit the proposed algorithm to $\mathbf{Y}_2$ under $\hat{\lambda}$
\end{algorithmic}
\end{algorithm}

\section{Simulation studies} \label{sec:4}

In this section, we present four simulation studies. Simulation study 1 evaluated the recovery of the sparse structure of factor loadings, the accuracy of parameter estimation, and 
%compared the estimation accuracy and 
computation time of the proposed method in moderate-scale settings. Since the MCMC estimation is also feasible at this scale, some measures were compared with those of the MCMC estimation. 
%between the BJME and MCMC methods. 
BJME finds estimates that maximize the complete-data joint posterior density based on the joint likelihood of the model parameters and factor scores. Accordingly, we compared the BJME estimates to the maximum a posteriori estimates obtained using the MCMC algorithm. 

Simulation study 2 compared the BJME estimates with those of MMLE under the moderate-scale settings. We considered two MMLE-based approaches. One is MMLE with the $L_1$-norm penalty, and the other is MMLE with rotation. The former aims to find the sparse latent structure by sparsely estimating factor loadings as in the proposed sparse BJME algorithm, while the latter explores the latent structure by rotation methods.

Simulation study 3 examined the performance of the BJME method in a large-scale and high-dimensional setting in which the MCMC method cannot be applied because the setting involves a large number of respondents, items, and latent factors. 

Simulation study 4 examined the performance of the BJME method when both the number of respondents and items increased simultaneously.

\subsection{Simulation study 1}

\subsubsection{Settings}
In Simulation study 1, we considered $N=\{500,1000\},J=30,K=3$, and $C_j=4$ for all $j$.
The true values $\nami{\boldsymbol{\theta}_i^*,\mathbf{a}_j^*,\mathbf{d}_j^*: i=1, \ldots, N, j=1,\ldots, J}$ were generated in the following manner. 
$\boldsymbol{\theta}_i^*$ was drawn from the multivariate normal distribution $\calN\maru{\mathbf{0}_K, \bSigma^*_\theta}$, where $\boldsymbol{\Sigma}_{\theta}^* = \rho \mathbf{1}_K\mathbf{1}_K^\top + (1-\rho)\mathbf{I}_K$ is the true covariance matrix of $\boldsymbol{\theta}^*$. $\rho$ represents the factor correlation, and we considered $\rho=\{0.1,0.4\}$ as the low and moderate correlations, respectively.
The factor loading parameter $\mathbf{a}_j^*$ was sparsely simulated as $\mathbf{u}_j^*\odot\mathbf{q}_j^*$, where $\mathbf{u}_j^*$ was generated from Uniform($0.5,2.0$), and $\mathbf{q}_j^*$ is the $j$-th row vector of the $J \times K$ binary matrix $\mathbf{Q}^*$ that represents the true sparse structure. We specified $\mathbf{Q}^*$ in such a manner that 60\%, 20\%, and 20\% of the items were loaded on one, two, and three latent factors, respectively. The intercept parameter $\mathbf{d}_j^*$ was generated to be strictly ordered: $d_{j1}^*$, $d_{j2}^*$, and $d_{j3}^*$ were drawn from Uniform($0.75, 1.5$), Uniform($-0.375,0.375$), and Uniform($-1.5,-0.75$), respectively. Fifty independent datasets were generated for each condition. 

For the proposed BJME method, we tuned the value of $\lambda$ by our CV procedure.
In addition to the proposed BJME algorithm, we also applied the MCMC estimation to the generated datasets. For both methods, we considered the true number of factors as known. These two methods shared the same prior settings in Equations~\ref{equ: theta prior}, \ref{equ: a prior}, and \ref{equ: d prior} for comparability, except for the penalty weight $\lambda$. For the MCMC method, we set the hyperprior over $\lambda$ to
\begin{align}
\lambda \sim \textrm{Half-Cauchy}(0,1).
\end{align}
We used the hyperprior over $\lambda$ instead of tuning $\lambda$ by the CV method because it is computationally infeasible to perform the CV procedure for the MCMC method. This is because our CV procedure requires an algorithm to be executed several times, whereas we cannot execute the MCMC algorithm many times due to its computational burden.
Two MCMC chains were used, with a burn-in period of 4000 iterations; the subsequent 4000 iterations were used to construct the posterior distributions of the parameters of interest. 
The convergence of each MCMC chain was verified by confirming that the potential scale reduction statistic was less than 1.1 \parencite[e.g.,][]{vehtari2021ranknormalization}. We performed the MCMC algorithm using the \texttt{cmdstanr} package \parencite{cmdstanr}, which is an interface to Stan \parencite{stan} in the R language \parencite{Rcoreteam}. The starting point for the MCMC algorithm was randomly generated from the same distributions as those of the true values. 

Moreover, we considered two initialization strategies for the proposed algorithm: a single start and multiple starts. Since the objective function in Equation~\ref{eq:objfun} we consider for the BJME method is non-concave, the proposed BJME algorithm can converge to a local optimum. To mitigate this issue, we compared estimation results obtained from a single start and multiple starts. For multiple starts, five different starting points were used, and we chose the estimate associated with the best objective function value among those. For the single start, we used the same initial values for both the BJME and MCMC methods.
All the starting values were generated from the same distributions as the true values.
Furthermore, we plugged in the covariance matrix of the latent factors, $\boldsymbol{\Sigma}_\theta$, by fitting a factor analysis based on the polychoric correlation matrix. We used the \texttt{fa} function of the \texttt{psych} package in R to compute the polychoric correlations and conduct a factor analysis. 

Due to the indeterminacy of column permutations and column sign flips in the estimated factor scores and factor loadings, it is possible to arbitrarily change the column permutations and column sign flips of the estimated factor loadings. This operation does not affect the resulting objective function value, provided that the corresponding permutations and sign flips are applied to the estimated factor scores.
Therefore, after obtaining the estimated factor loadings, we changed the column permutations and column sign flips of $\hat{\mathbf{A}}$ such that $\sum_{j=1}^J\|\hat{\mathbf{a}}_j-\mathbf{a}_j^*\|$ is minimized using the \texttt{matchICA} function of the \texttt{steadyICA} package. 

We examined the effectiveness of the regularization method in distinguishing between zero and non-zero factor loadings. For this purpose, we calculated the misselection rate (MSR), false positive rate (FPR), and false negative rate (FNR) to evaluate the accuracy of variable selection over latent factors:
\begin{align}
\textrm{MSR}(\hat{\mathbf{Q}},\mathbf{Q}^*) &= \frac{\sum_{j=1}^J\sum_{k=1}^K\mathbb{1}\{\hat{q}_{jk}\neq q_{jk}^*\}}{JK}, \\
\textrm{FPR}(\hat{\mathbf{Q}},\mathbf{Q}^*) &= \frac{\sum_{j=1}^J\sum_{k=1}^K\mathbb{1}\{\hat{q}_{jk}=1,q_{jk}^*=0\}}{\sum_{j=1}^J\sum_{k=1}^K\mathbb{1}\{q_{jk}^*=0\}}, \\
\textrm{FNR}(\hat{\mathbf{Q}},\mathbf{Q}^*) &= \frac{\sum_{j=1}^J\sum_{k=1}^K\mathbb{1}\{\hat{q}_{jk}=0,q_{jk}^*=1\}}{\sum_{j=1}^J\sum_{k=1}^K\mathbb{1}\{q_{jk}^*=1\}},
\end{align}
where $\hat{\mathbf{Q}} = (\hat{q}_{jk})_{J\times K}$ is an estimated sparse structure. For the proposed algorithm, we computed $\hat{q}_{jk} = \mathbb{1}\{|\hat{a}_{jk}|>0.01\}$. 
For the MCMC, we considered $\hat{q}_{jk} = 1$ when the 95\% credible interval of $a_{jk}$ excluded zero and $\hat{q}_{jk} = 0$ otherwise.

Moreover, to measure estimation accuracy, we investigated the recovery error of the model parameters by computing the following quantities:
\begin{align}
\textrm{Error}(\hat{\mathbf{A}},\mathbf{A}^*) &= \left(\frac{1}{\sum_{j=1}^J\sum_{k=1}^K \mathbb{1}\{q_{jk}^*=1\}} \sum_{j=1}^J\|\mathbf{q}_j^*\odot(\hat{\mathbf{a}}_j - \mathbf{a}_j^*)\|^2 \right)^{1/2}, \\
\textrm{Error}(\hat{\mathbf{D}},\mathbf{D}^*) &= \left(\frac{1}{J}\sum_{j=1}^J\frac{\|\hat{\mathbf{d}}_j - \mathbf{d}_j^*\|^2}{C_j-1} \right)^{1/2}, \\
\textrm{RelativeBias}(\hat{\mathbf{A}},\mathbf{A}^*) &= \frac{1}{\sum_{j=1}^J\sum_{k=1}^K \mathbb{1}\{q_{jk}^*=1\}} \sum_{j=1}^J \sum_{k=1}^K q_{jk}^* \frac{\hat{a}_{jk} - a^*_{jk}}{a^*_{jk}}, \\
\textrm{RelativeBias}(\hat{\mathbf{D}},\mathbf{D}^*) &= \frac{1}{J} \sum_{j=1}^J \frac{1}{C_j-1} \sum_{c=1}^{C_j-1} \frac{\hat{d}_{jc} - d^*_{jc}}{d^*_{jc}},
\end{align}
Regarding the error metrics of factor loadings, we computed these values in the non-zero entries of the true loadings.
% We then create box plots of these values across 50 replications. 
Furthermore, we evaluated the computation time to assess the computational efficiency. To perform BJME and MCMC, we used two cores for parallel computing. The analyses were conducted on a desktop computer with AMD Ryzen Threadripper 3970X 32-core Processor 3.69 GHz and 256 GB RAM.

\subsubsection{Results}
\begin{figure}[htp]% [H] is so declass\'e!
\begin{minipage}{0.3cm}
\rotatebox{90}{Rate}
\end{minipage}\hfill
\centering
\begin{minipage}{0.45\textwidth}
\centering
$N=500,\rho=0.1$\\
\includegraphics[width=\textwidth]{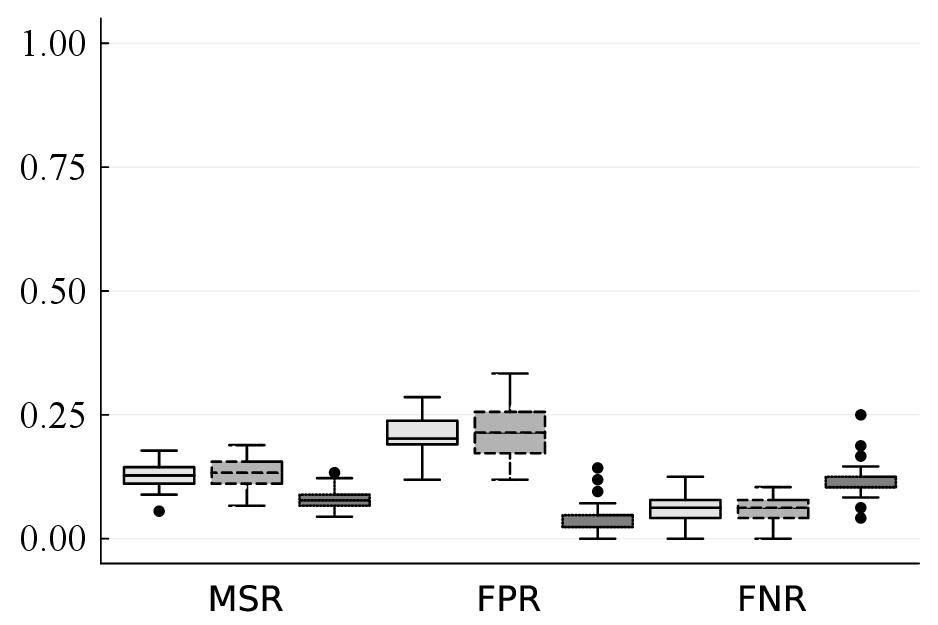}
\end{minipage}\hfill
\begin{minipage}{0.45\textwidth}
\centering
$N=500,\rho=0.4$\\
\includegraphics[width=\textwidth]{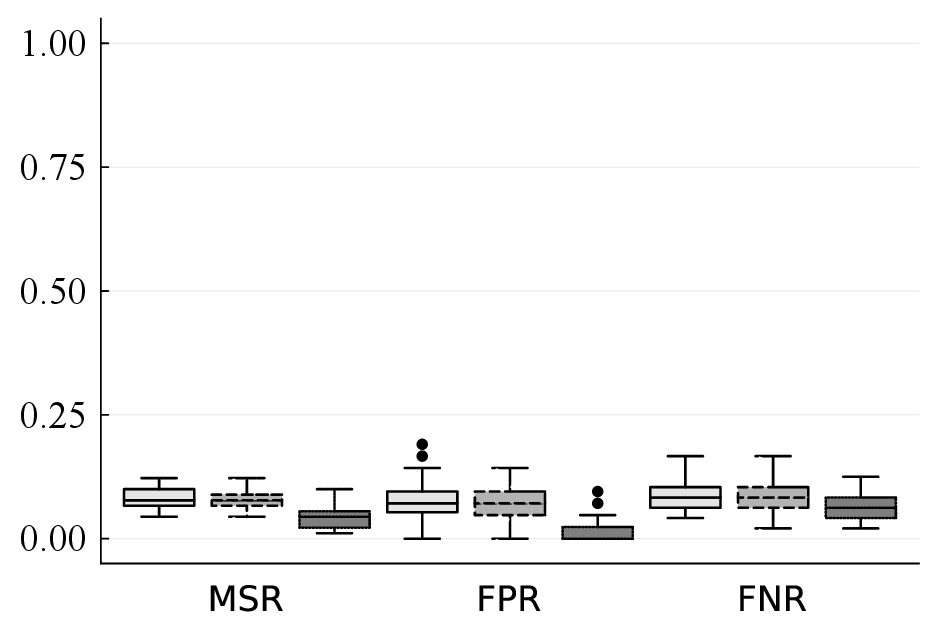}
\end{minipage}\\
\begin{minipage}{0.3cm}
\rotatebox{90}{Rate}
\end{minipage}\hfill
\centering
\begin{minipage}{0.45\textwidth}
\centering
$N=1000,\rho=0.1$\\
\includegraphics[width=\textwidth]{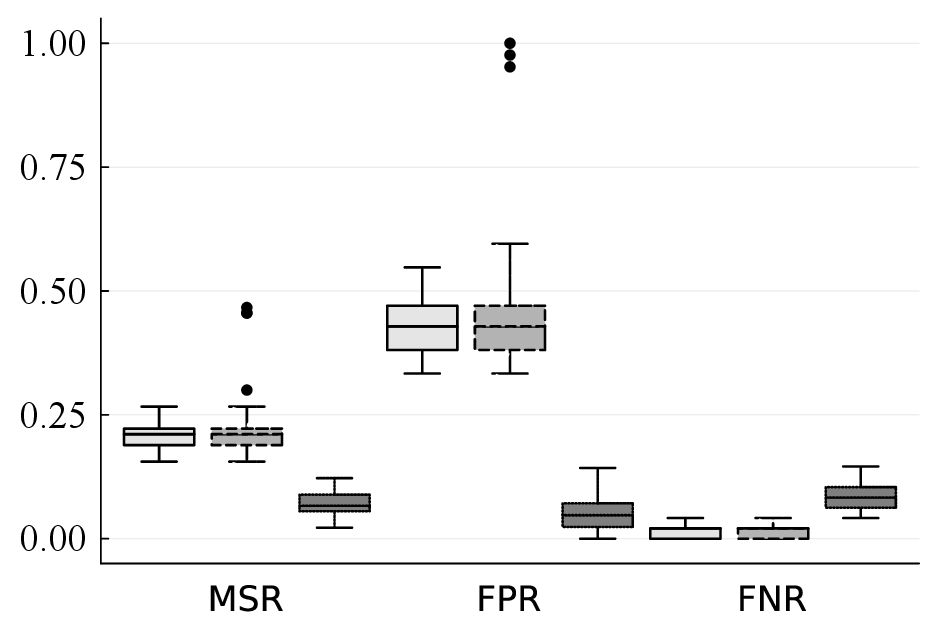}
\end{minipage}\hfill
\begin{minipage}{0.45\textwidth}
\centering
$N=1000,\rho=0.4$\\
\includegraphics[width=\textwidth]{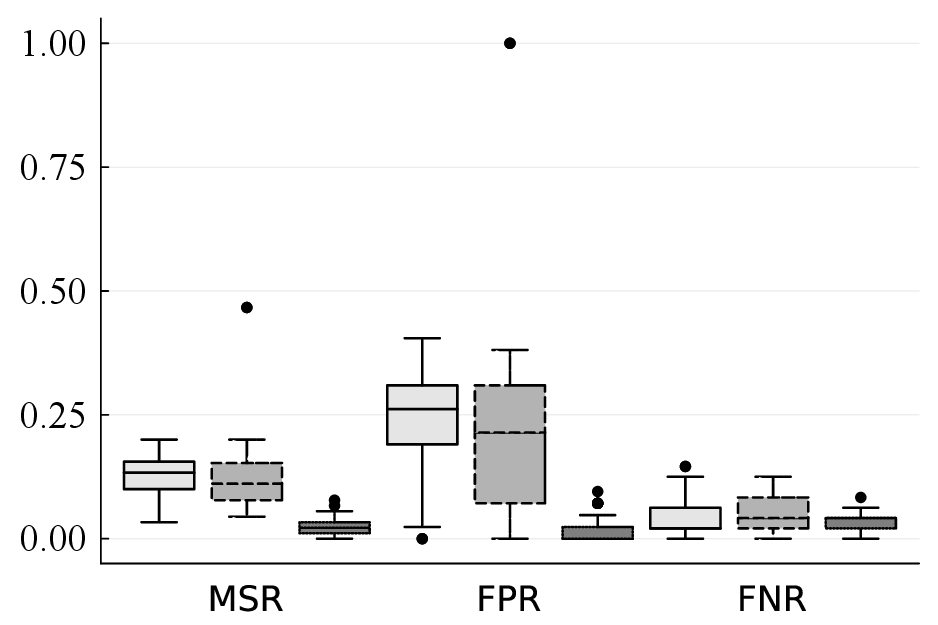}
\end{minipage}\\
\includegraphics[width=0.8\textwidth]{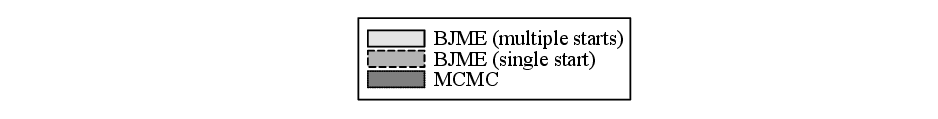}
\caption{Selection error in Simulation study 1}\label{fig: sim1 select error}
\end{figure}

Figure~\ref{fig: sim1 select error} shows the accuracy of variable selection over latent factors in terms of MSR, FPR, and FNR.
The MCMC method was slightly superior to the BJME method in MSR under all conditions, which indicates that the MCMC method is slightly more accurate in recovering the true sparse structure than the BJME method.
The resulting values of the MSR and FNR of the BJME method were also small, indicating the good performance of the proposed BJME in variable selection. The FNR of the BJME method was very small compared to the FPR in both $N=500$ and $N=1000$. These results imply that our CV procedure tends to select a relatively small value of $\lambda$.

Figure~\ref{fig: sim1 select error} also illustrates the impact of the factor correlation on the accuracy of variable selection. For both BJME and MCMC, the selection accuracy slightly improved when the factor correlation increased from 
$0.1$ to $0.4$. For BJME, a higher factor correlation led to a slight decrease in the MSR and the FPR, while the FNR increased slightly. In contrast, for MCMC, all three error measures decreased when the factor correlation became larger. Overall, these results suggest that both methods maintain stable selection performance even when the latent factors are moderately correlated.

\begin{figure}[htp]% [H] is so declass\'e!
\begin{minipage}{0.3cm}
\rotatebox{90}{Error}
\end{minipage}\hfill
\centering
\begin{minipage}{0.4\textwidth}
\centering
$N=500,\rho=0.1$\\
\includegraphics[width=\textwidth]{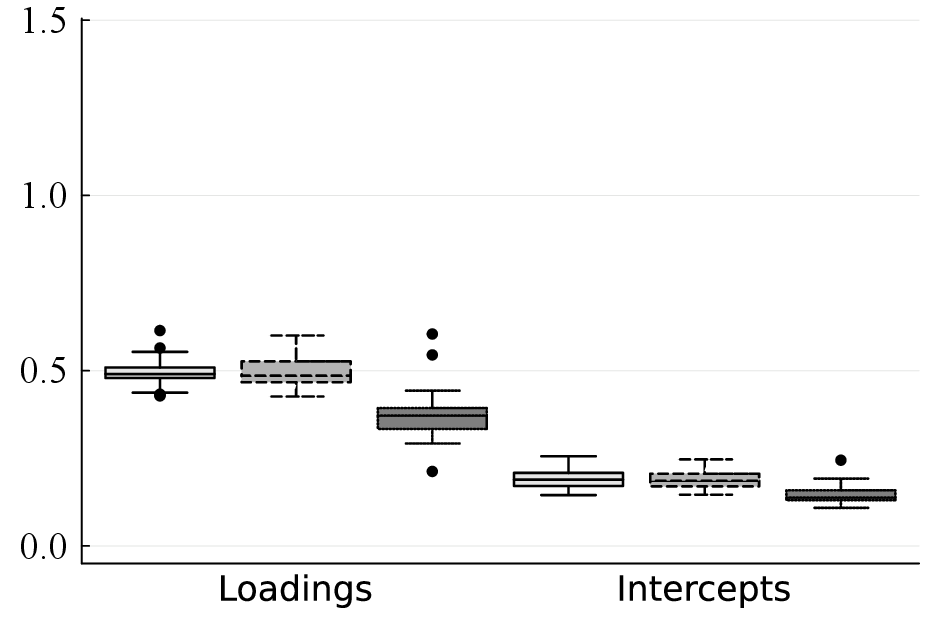}
\end{minipage}\hfill
\begin{minipage}{0.4\textwidth}
\centering
$N=500,\rho=0.4$\\
\includegraphics[width=\textwidth]{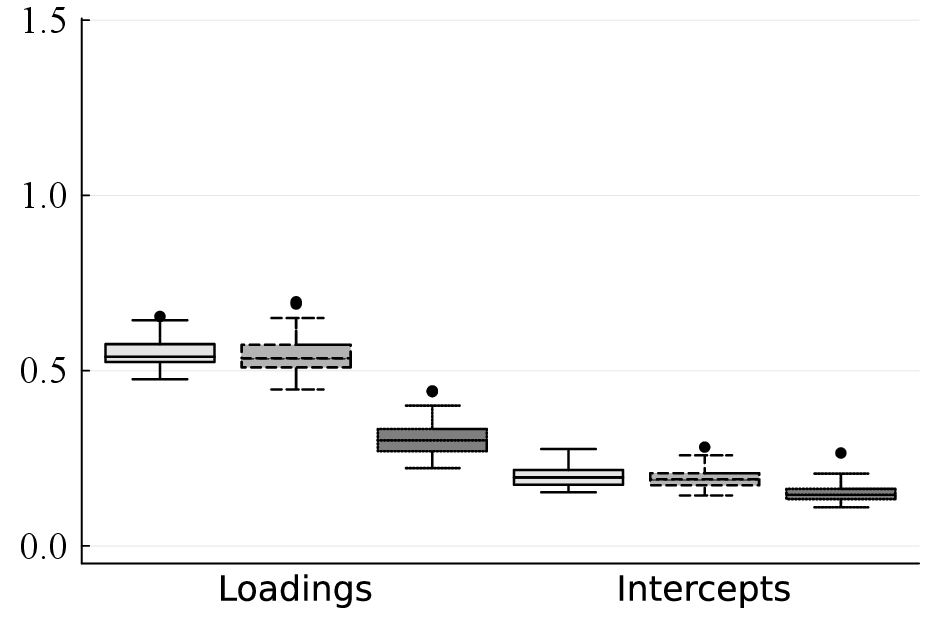}
\end{minipage}\\
\begin{minipage}{0.3cm}
\rotatebox{90}{Error}
\end{minipage}\hfill
\centering
\begin{minipage}{0.4\textwidth}
\centering
$N=1000,\rho=0.1$\\
\includegraphics[width=\textwidth]{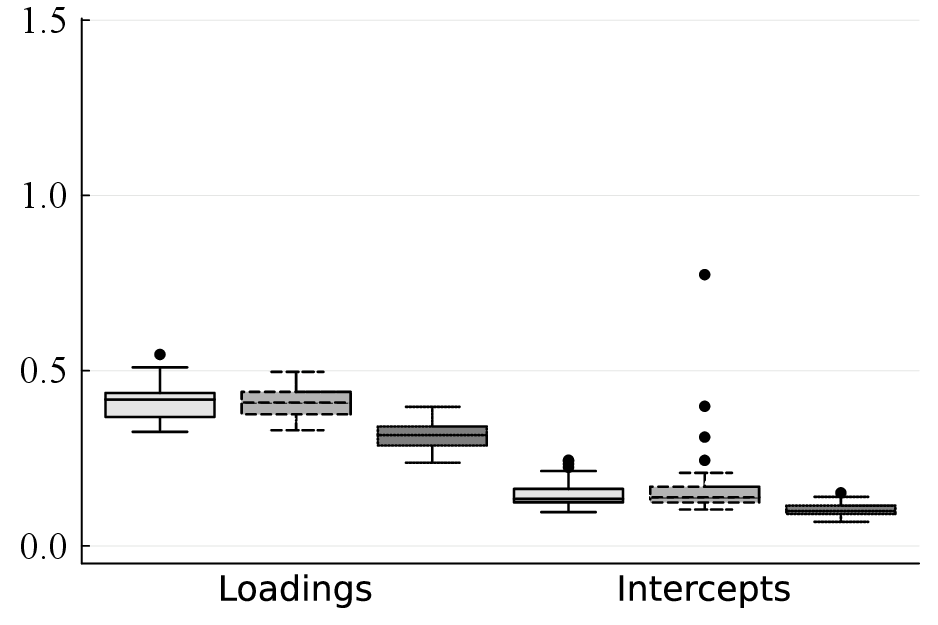}
\end{minipage}\hfill
\begin{minipage}{0.4\textwidth}
\centering
$N=1000,\rho=0.4$\\
\includegraphics[width=\textwidth]{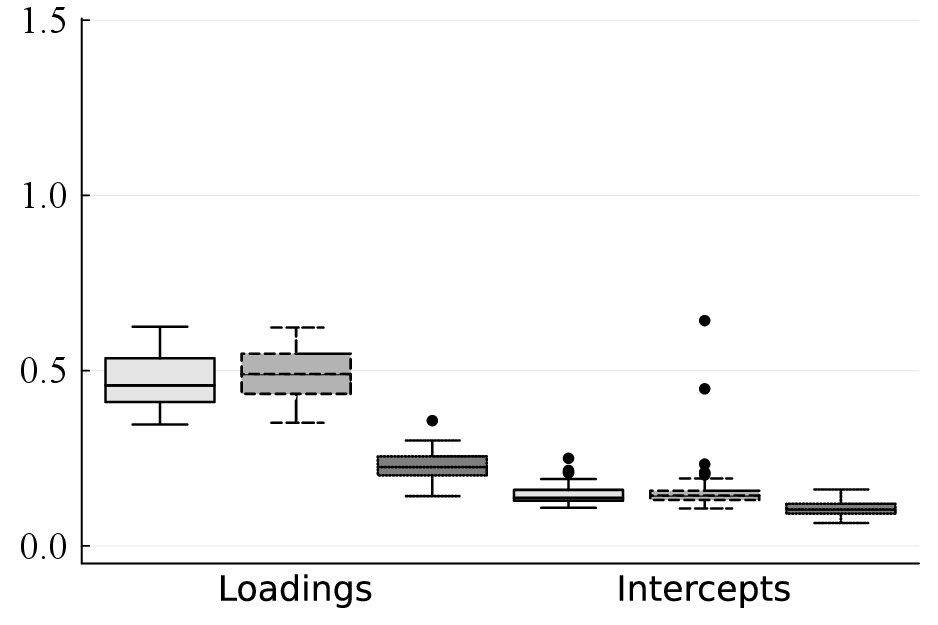}
\end{minipage}\\
\begin{minipage}{0.3cm}
\rotatebox{90}{Relative bias}
\end{minipage}\hfill
\centering
\begin{minipage}{0.4\textwidth}
\centering
$N=500,\rho=0.1$\\
\includegraphics[width=\textwidth]{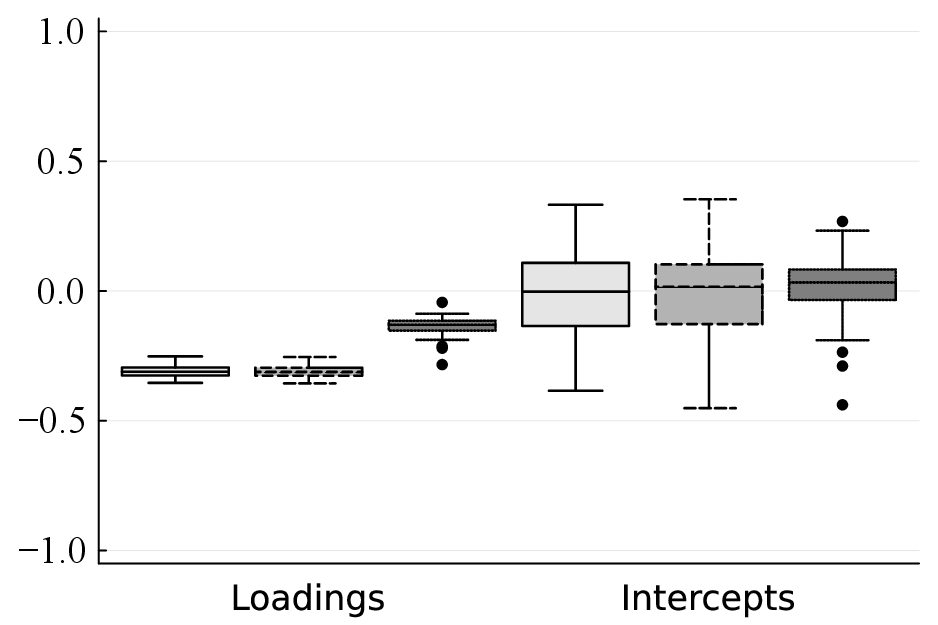}
\end{minipage}\hfill
\begin{minipage}{0.4\textwidth}
\centering
$N=500,\rho=0.4$\\
\includegraphics[width=\textwidth]{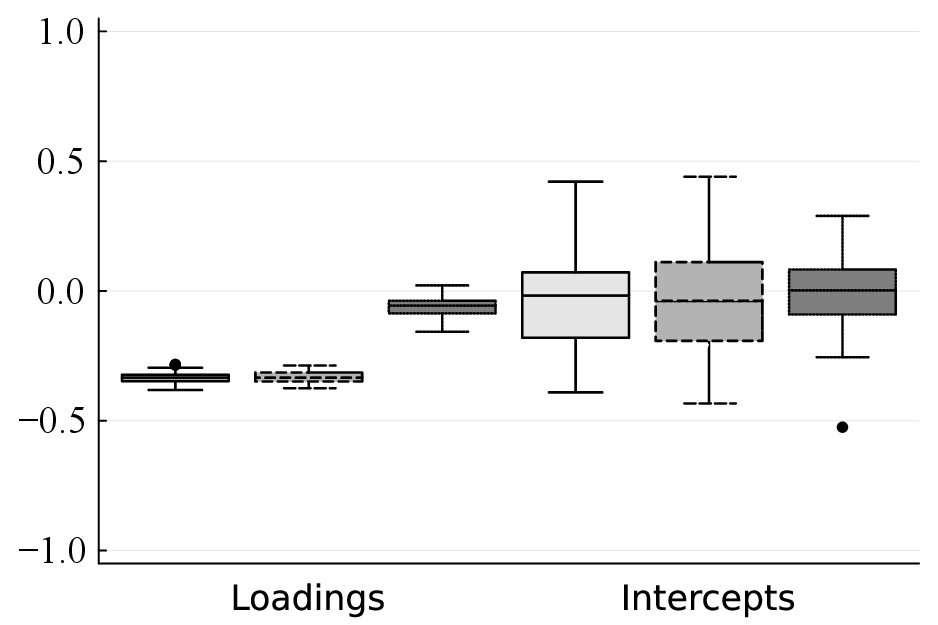}
\end{minipage}\par
\begin{minipage}{0.3cm}
\rotatebox{90}{Relative bias}
\end{minipage}\hfill
\centering
\begin{minipage}{0.4\textwidth}
\centering
$N=1000,\rho=0.1$\\
\includegraphics[width=\textwidth]{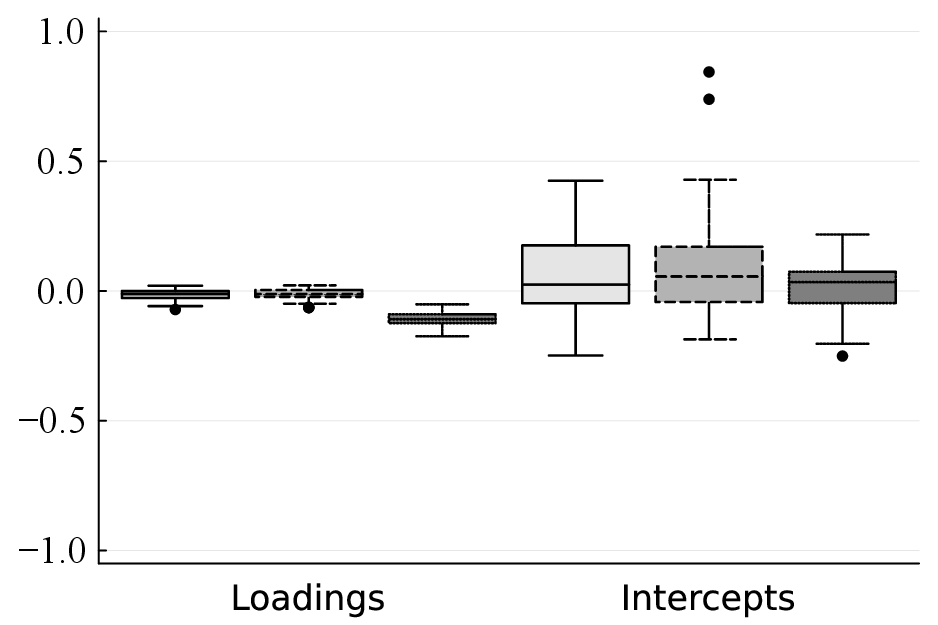}
\end{minipage}\hfill
\begin{minipage}{0.4\textwidth}
\centering
$N=1000,\rho=0.4$\\
\includegraphics[width=\textwidth]{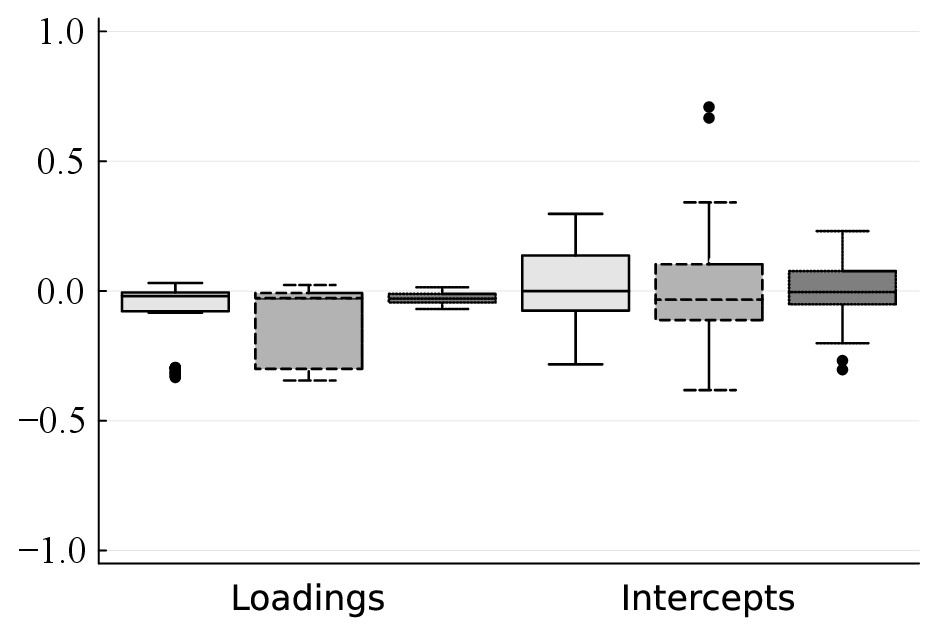}
\end{minipage}\\
\includegraphics[width=0.8\textwidth]{images_v2/sim1_legend.eps}
\caption{Recovery error of the model parameters in Simulation study 1}\label{fig: sim1 rmse}
\end{figure}

Next, the estimation accuracy of the BJME is examined in comparison with the MCMC method.
The box plots of the errors are presented in Figure~\ref{fig: sim1 rmse}.
For all methods, the estimates in $N=1000$ were more accurate than in $N=500$, which indicates that increasing the number of respondents improves the estimation accuracy. The MCMC method outperformed the BJME method with a single start, indicating that the MCMC method can achieve more accurate estimates than the BJME method in $N=500$ and $N=1000$ despite the same starting point.
Regarding the relative bias of factor loadings, the MCMC method was superior to the BJME methods under $N=500$. Increasing the sample size also improved the bias of factor loadings, especially for the BJME methods. For all methods, the intercept estimates were less biased than those of factor loadings, which would be the effect of the small penalty weight by the large-variance normal prior $\sigma_d^2=100^2$.

We also examined the impact of factor correlation on the estimation accuracy of the model parameters. Overall, both BJME and MCMC showed little sensitivity to the correlation level, as the recovery error remained largely unchanged when $\rho$ increased from $0.1$ to $0.4$. This suggests that both methods are robust to moderate correlations among latent factors in terms of parameter recovery. One exception was observed for the relative bias of the loadings under BJME, where the variability becomes slightly larger at $\rho=0.4$, as indicated by a wider spread in the boxplots. However, this effect is limited and does not substantially affect the overall estimation accuracy.

The average computation time and selected $\lambda$ values over 50 simulated datasets are listed in Table~\ref{tab:time simulation1}.
\begin{table}[htbp]
\caption{Average computation time (seconds) and selected $\lambda$ values over 50 replications in Simulation study 1}
\centering
\begin{tabular}{lcccc}
\hline
 & \multicolumn{2}{c}{$N=500$} & \multicolumn{2}{c}{$N=1000$} \\
 \cline{2-5}
 & $\rho=0.1$ & $\rho=0.4$ & $\rho=0.1$ & $\rho=0.4$ \\ 
\hline
BJME (multiple starts) 
& 57.83 (14.00) & 75.02 (14.00) & 268.27 (14.00) & 323.40 (16.88) \\
BJME (single start) 
& 13.22 (14.00) & 13.92 (14.00) & 40.50 (13.17) & 41.62 (17.76) \\
MCMC 
& 4271.06 & 4523.28 & 8213.86 & 9125.59 \\
\hline
\end{tabular}
\label{tab:time simulation1}
\end{table}
Here, values in parentheses indicate the average selected $\lambda$.
The computation time for the BJME corresponds to the time elapsed in our CV procedure.
The computation time of the MCMC method is the time required for both chains to finish 8000 iterations. Table~\ref{tab:time simulation1} shows that the BJME method is much more computationally efficient than the MCMC method.

\subsection{Simulation study 2}

\subsubsection{Settings}

In Simulation study 2, we compared the performance of the proposed sparse BJME method with two alternative approaches based on MMLE: the sparse MMLE and MMLE with rotation. To perform MMLE, we employed the \texttt{sa\_penmirt} function of the \texttt{lvmcomp} package, which is written in C++. The function implements a stochastic approximation EM algorithm for MMLE in IFA, which was developed in \textcite{zhang2020improved}. We adopted the quartimin rotation as the rotation method.

Regarding the data generation process, we considered dichotomous responses, i.e., $C_j=2$ for all $j$, because the \texttt{sa\_penmirt} function does not support polytomous responses. In this setting, the MGRM is reduced to the multidimensional two-parameter logistic (M2PL) model, where the response probabilities are modeled as
\begin{align}
P(Y_{ij}=0 \mid \boldsymbol{\theta}_i,\mathbf{a}_j,d_j) &= 1 - P(Y_{ij}=1 \mid \boldsymbol{\theta}_i,\mathbf{a}_j,d_j), \\
P(Y_{ij}=1 \mid \boldsymbol{\theta}_i,\mathbf{a}_j,d_j) &= \textrm{logit}^{-1}(\boldsymbol{\theta}_i^\top\mathbf{a}_j+d_j),
\end{align}
where $d_j \in \mathbb{R}$ is the intercept parameter for item $j$. The number of intercepts of the M2PL model is different from that of the MGRM, and we generate the true values $d_j^*$ from Uniform($-1.5, 1.5$) for all $j$.
The other settings, including the number of respondents and items, factor correlation, and the design of the true Q-matrix, were the same as in Simulation study 1.

For the sparse MMLE, we tuned the penalty weight using the Bayesian information criterion (BIC):
% \begin{align}
% \textrm{BIC}(\lambda) = -2\sum_{i=1}^N \log \left(\int_{\mathbb{R}^K} \prod_{j=1}^J P(Y_{ij}, \boldsymbol{\theta}_i \mid \mathbf{a}_j,d_j ; \mathbf{0}_K,\boldsymbol{\Sigma}_\theta) d\boldsymbol{\theta}_i \right) + \log (N) \sum_{j=1}^J \|\mathbf{a}_j\|_0,
% \end{align}
% equation below was added by Okada
\begin{align}
\textrm{BIC}(\lambda)
= -2\sum_{i=1}^N
\log \left[\int_{\mathbb{R}^K} \left\{ \prod_{j=1}^J P(Y_{ij}\mid \boldsymbol{\theta}_i,\mathbf{a}_j,d_j)
\right\} P(\boldsymbol{\theta}_i\mid \mathbf{0}_K,\boldsymbol{\Sigma}_\theta) d\boldsymbol{\theta}_i \right]
+ \log(N)\sum_{j=1}^J \|\mathbf{a}_j\|_0.
\end{align}
where $\|\cdot\|_0$ is the $L_0$ quasi-norm, which counts the number of nonzero elements. This is because existing studies involving the sparse MMLE \parencite{sun2016latent,cho2022regularized} determine the penalty weight according to information criteria. It should be noted that we approximated the log marginal likelihood in the BIC using the Gauss--Hermite quadrature. The selection process for the sparse MMLE followed a data-splitting and two-stage selection strategy as in our CV procedure. First, we randomly divided the dataset into training and test sets. We then fitted the sparse MMLE algorithm to the training set using an initial set of candidate penalty weights, $\lambda^{(1)}=\{0.01, 0.1, 1, 10, 100\}$. Based on the BIC, we identified the optimal value $\hat{\lambda}^{(1)}$ from this first stage. In the second stage, we constructed a refined set of candidate values $\lambda^{(2)}$ centered around $\hat{\lambda}^{(1)}$, and refitted the algorithm to the training set. Finally, we applied the selected penalty weight $\hat{\lambda}$, the second-stage minimizer of BIC, to the test set. For the sparse MMLE, we truncated the estimated factor loadings whose absolute values were less than 0.01, as in the proposed algorithm, while we truncated the rotated factor loadings to zero if their absolute values were less than 0.1 for the MMLE with rotation to enhance sparsity.

Regarding the starting point, a single start was employed because we observed a negligible difference in terms of the accuracy of parameter recovery and variable selection over latent factors between a single start and multiple starts in Simulation study 1. The BJME and MMLE shared the same starting point for comparability. Since the \texttt{sa\_penmirt} function does not support parallel computing, we performed the BJME and MMLE using a single core on the same computing platform as that used in Simulation study 1. Moreover, we ran the stochastic approximation EM algorithm implemented in \texttt{sa\_penmirt} for 1000 iterations because fewer iterations, such as 200, did not achieve good estimation results.

\subsubsection{Results}

\begin{figure}[htp]% [H] is so declass\'e!
\begin{minipage}{0.08\linewidth}
\rotatebox{90}{Rate}
\end{minipage}\hfill
\centering
\begin{minipage}{0.45\textwidth}
\centering
$N=500,\rho=0.1$\\
\includegraphics[width=\textwidth]{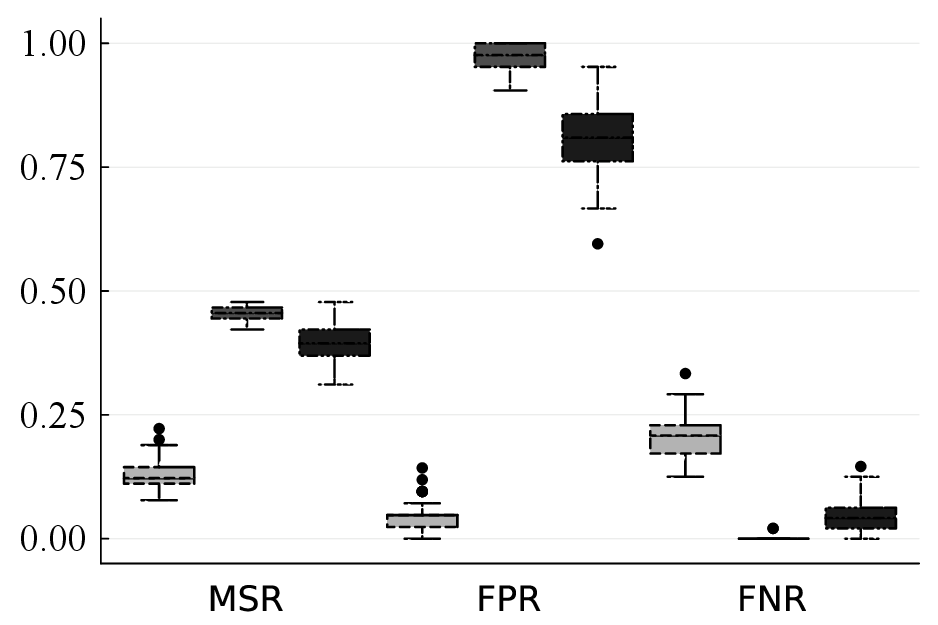}
\end{minipage}\hfill
\begin{minipage}{0.45\textwidth}
\centering
$N=500,\rho=0.4$\\
\includegraphics[width=\textwidth]{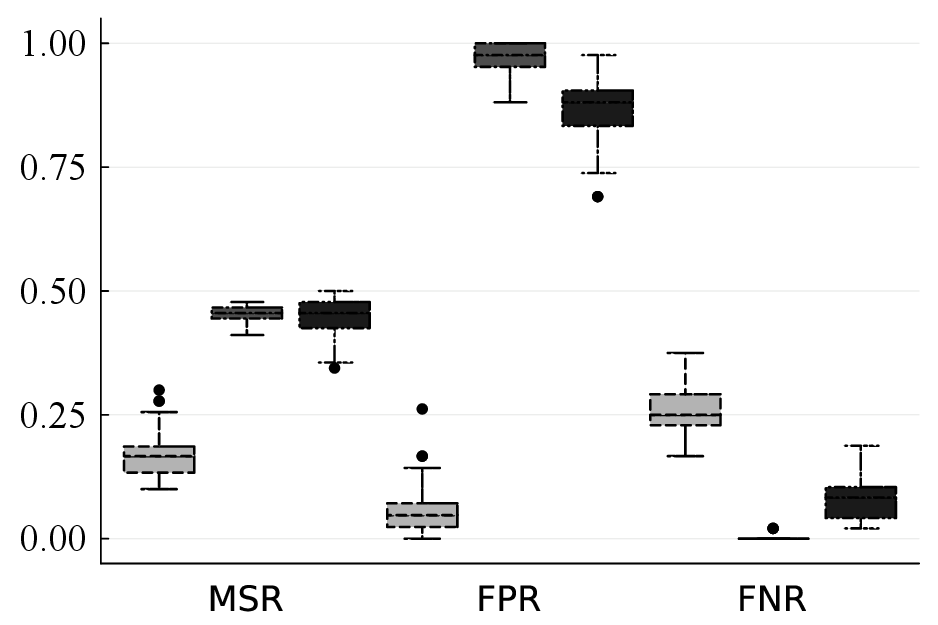}
\end{minipage}\\
\begin{minipage}{0.08\linewidth}
\rotatebox{90}{Rate}
\end{minipage}\hfill
\centering
\begin{minipage}{0.45\textwidth}
\centering
$N=1000,\rho=0.1$\\
\includegraphics[width=\textwidth]{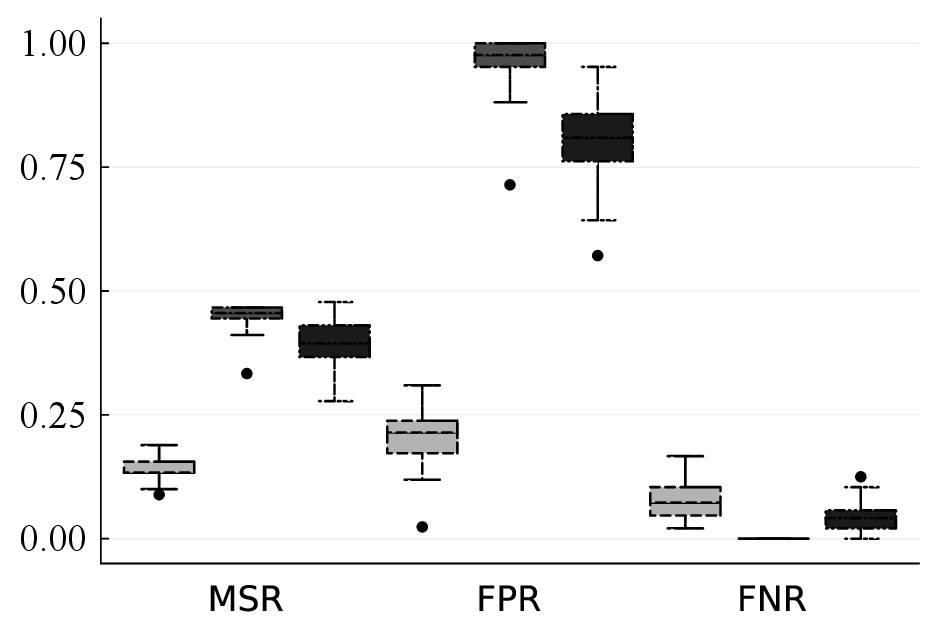}
\end{minipage}\hfill
\begin{minipage}{0.45\textwidth}
\centering
$N=1000,\rho=0.4$\\
\includegraphics[width=\textwidth]{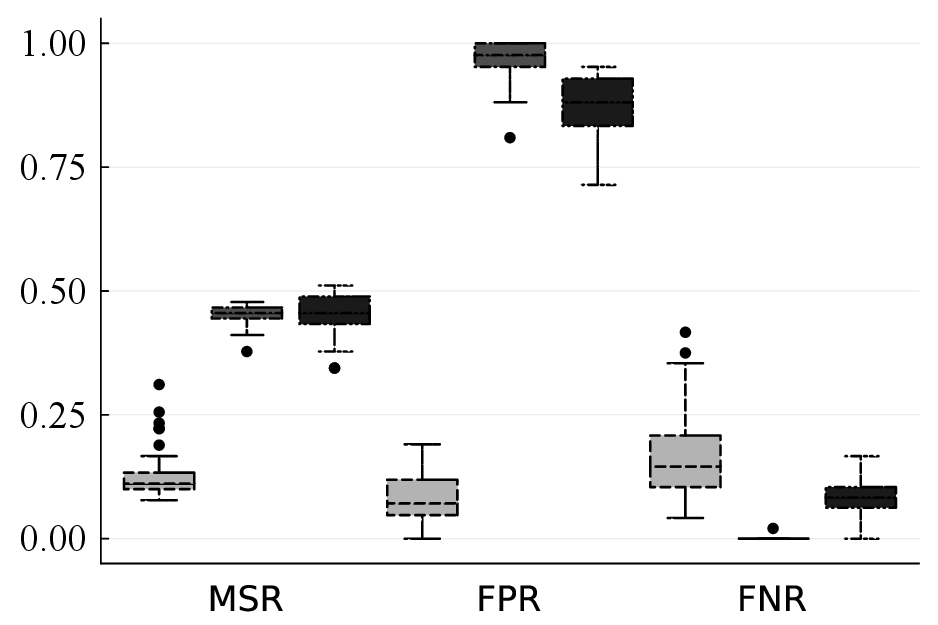}
\end{minipage}\\
\includegraphics[width=0.8\textwidth]{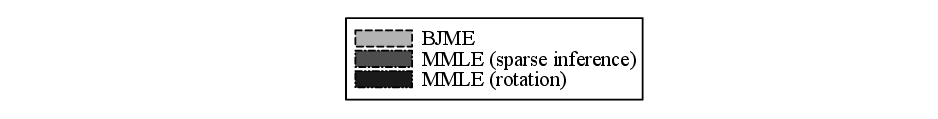}
\caption{Selection error in Simulation study 2}\label{fig: sim2 select error}
\end{figure}

The box plots of the selection error are presented in Figure~\ref{fig: sim2 select error}.
The proposed BJME method demonstrated superior performance in terms of MSR, compared to both sparse MMLE and MMLE with rotation, which indicates that the BJME method is more accurate in latent variable selection than the MMLE-based approaches.
The FPR and FNR of the MMLE were quite high and low, respectively. This shows that the sparse MMLE struggles to effectively tune $\lambda$ via our two-stage selection strategy and the BIC. Furthermore, the MMLE with rotation performs poorly in achieving sparsity, although we truncated the rotated loadings to zero if their absolute values are less than 0.1.
This finding supports our initial assertion that the regularization approach inherent in the proposed BJME algorithm is more effective at shrinking truly irrelevant factor loadings to zero, thus correctly identifying more zero loadings.

\begin{figure}[htp]% [H] is so declass\'e!
\begin{minipage}{0.08\linewidth}
\rotatebox{90}{Error}
\end{minipage}\hfill
\centering
\begin{minipage}{0.4\textwidth}
\centering
$N=500,\rho=0.1$\\
\includegraphics[width=\textwidth]{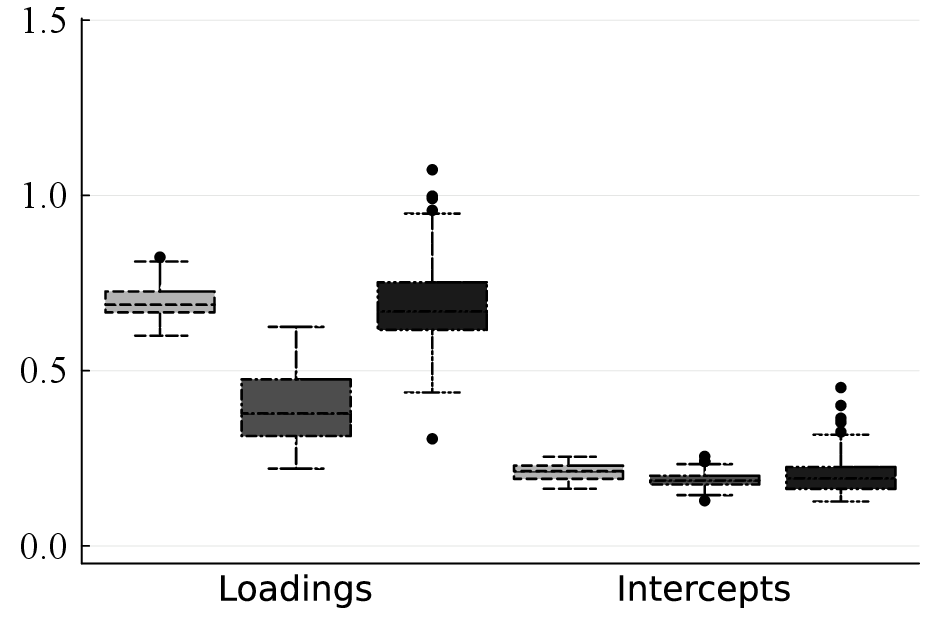}
\end{minipage}\hfill
\begin{minipage}{0.4\textwidth}
\centering
$N=500,\rho=0.4$\\
\includegraphics[width=\textwidth]{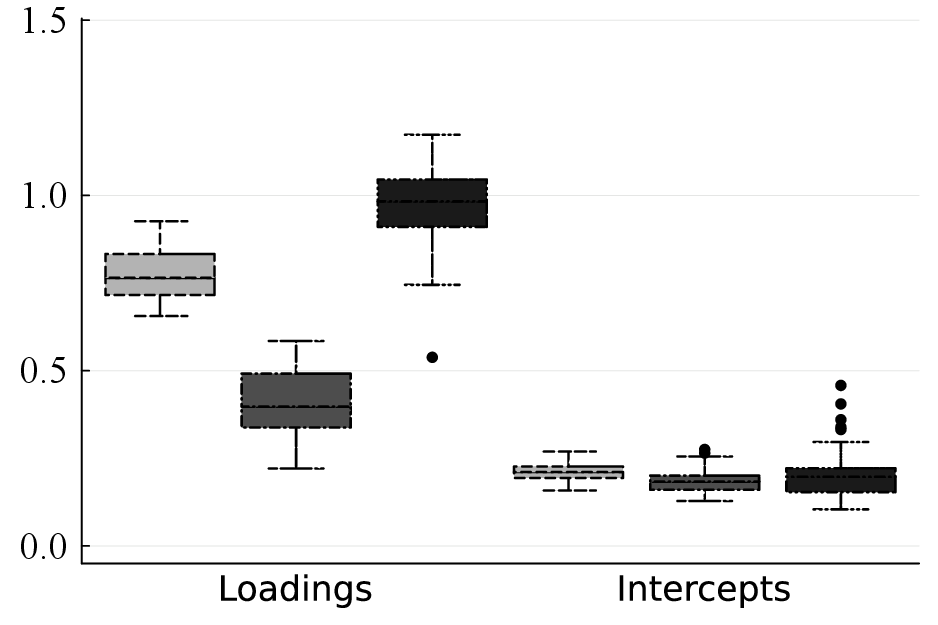}
\end{minipage}\\
\begin{minipage}{0.08\linewidth}
\rotatebox{90}{Error}
\end{minipage}\hfill
\centering
\begin{minipage}{0.4\textwidth}
\centering
$N=1000,\rho=0.1$\\
\includegraphics[width=\textwidth]{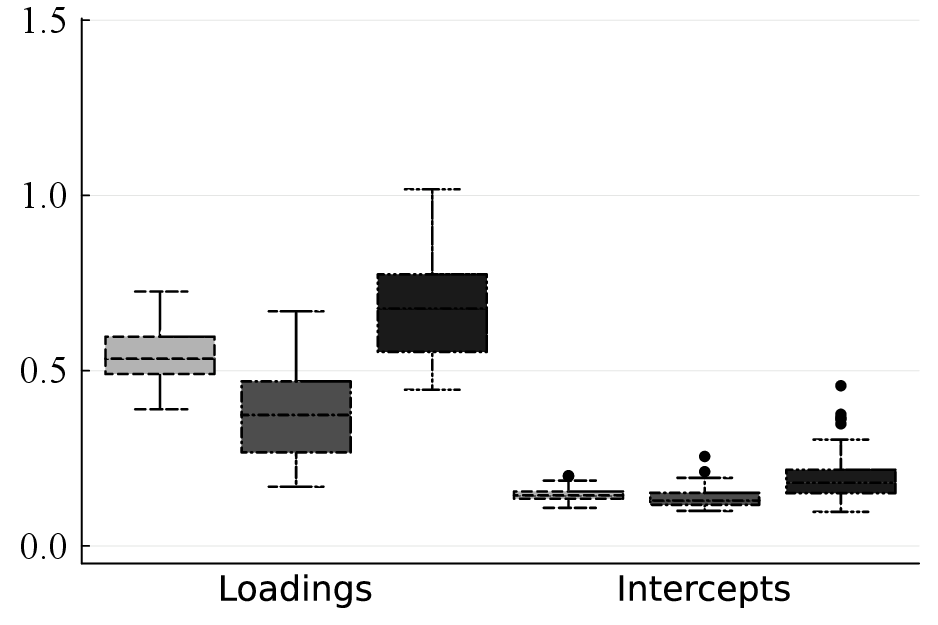}
\end{minipage}\hfill
\begin{minipage}{0.4\textwidth}
\centering
$N=1000,\rho=0.4$\\
\includegraphics[width=\textwidth]{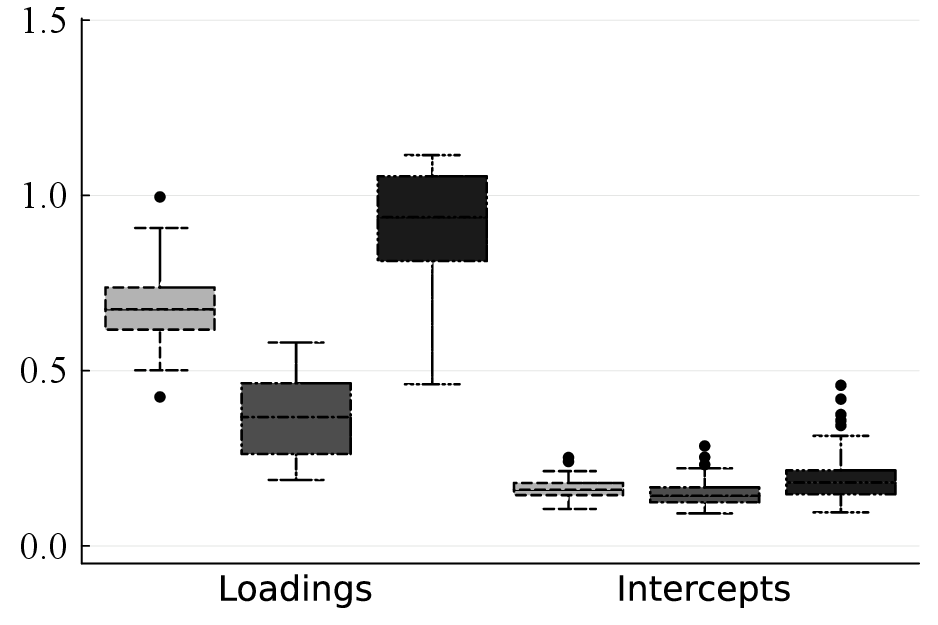}
\end{minipage}\\
\begin{minipage}{0.08\linewidth}
\rotatebox{90}{Relative bias}
\end{minipage}\hfill
\centering
\begin{minipage}{0.4\textwidth}
\centering
$N=500,\rho=0.1$\\
\includegraphics[width=\textwidth]{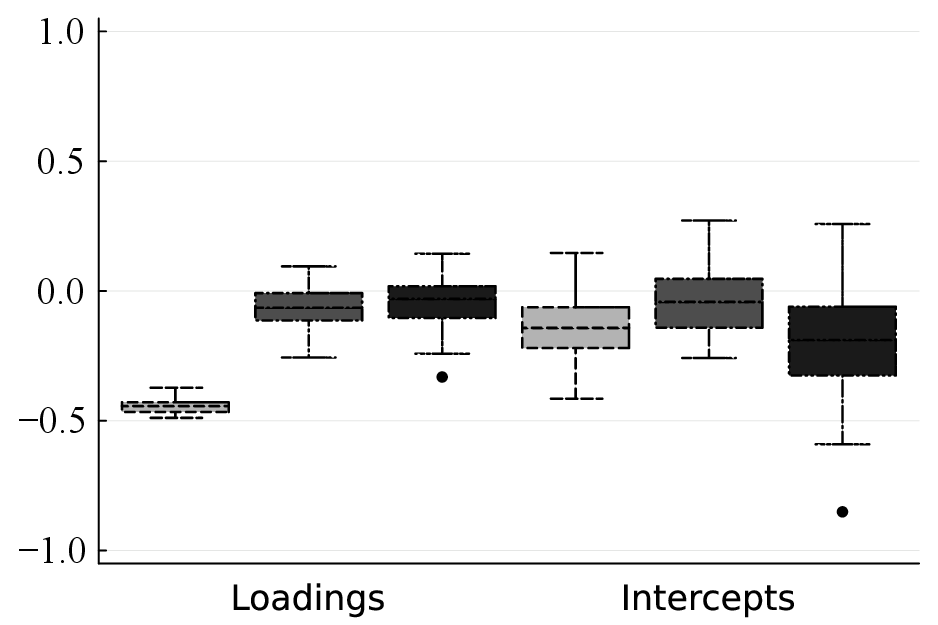}
\end{minipage}\hfill
\begin{minipage}{0.4\textwidth}
\centering
$N=500,\rho=0.4$\\
\includegraphics[width=\textwidth]{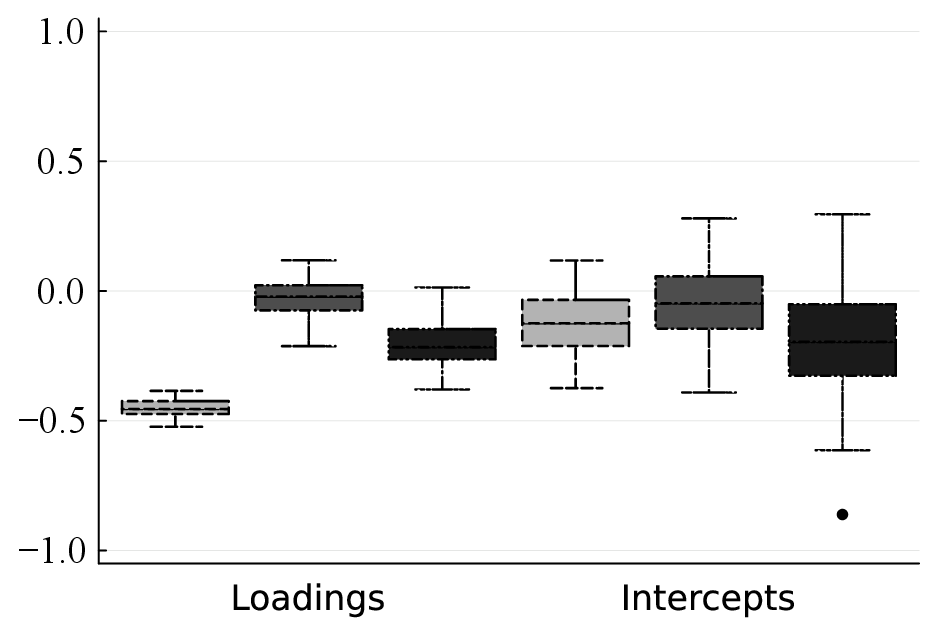}
\end{minipage}\\
\begin{minipage}{0.08\linewidth}
\rotatebox{90}{Relative bias}
\end{minipage}\hfill
\centering
\begin{minipage}{0.4\textwidth}
\centering
$N=1000,\rho=0.1$\\
\includegraphics[width=\textwidth]{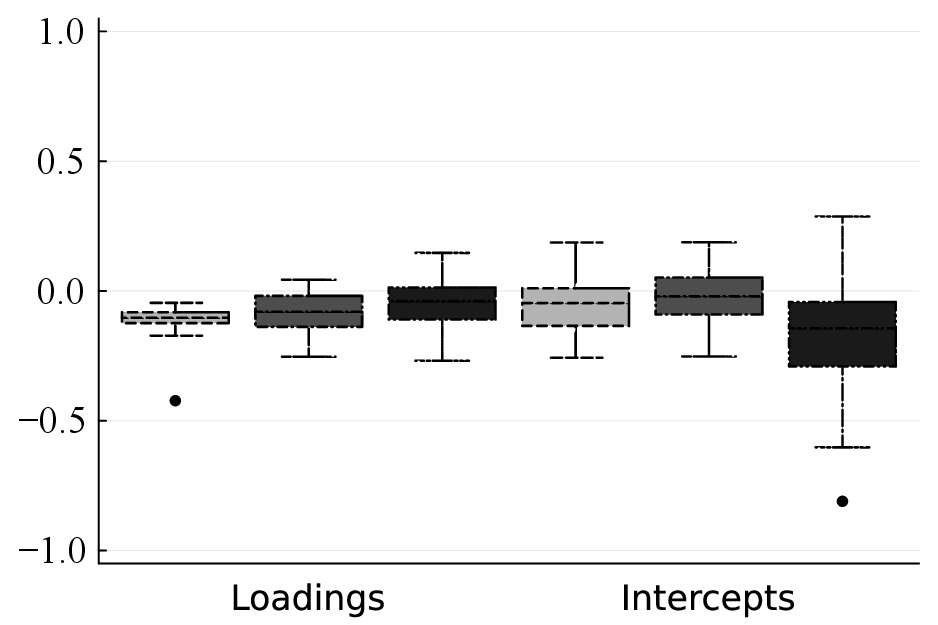}
\end{minipage}\hfill
\begin{minipage}{0.4\textwidth}
\centering
$N=1000,\rho=0.4$\\
\includegraphics[width=\textwidth]{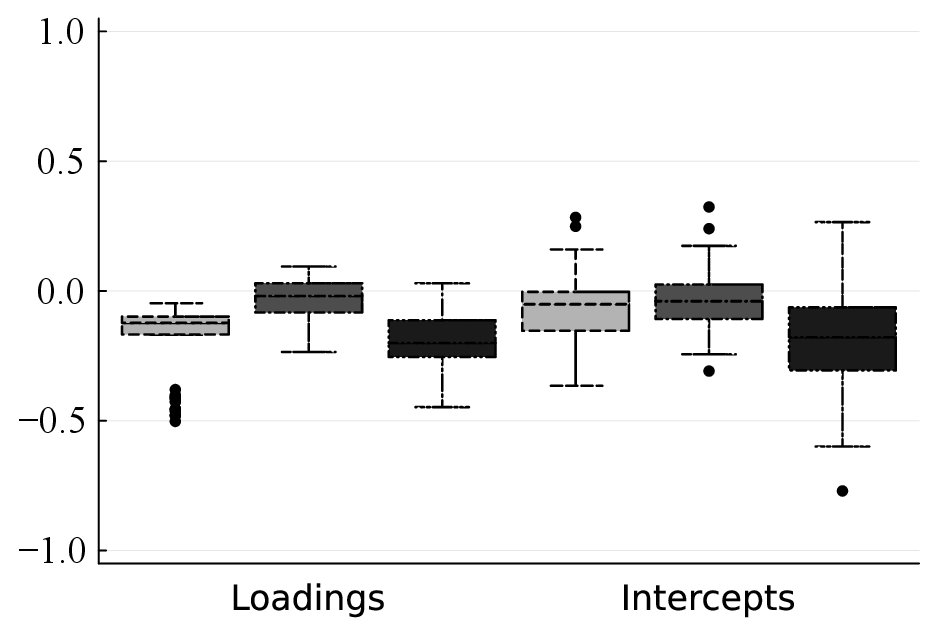}
\end{minipage}\\
\includegraphics[width=0.8\textwidth]{images_v2/sim2_legend.eps}
\caption{Recovery error of the model parameters in Simulation study 2}\label{fig: sim2 rmse}
\end{figure}

Then, we compare the estimation accuracy. The box plots of the error metric values are presented in Figure~\ref{fig: sim2 rmse}.
The sparse MMLE method showed smaller error and bias
than the BJME method, which is consistent with the result of Simulation study 1 because the likelihood function considered in the MCMC algorithm is the marginal likelihood. Furthermore, this trend is also consistent with existing studies that compared JMLE and MMLE \parencite{chen2019joint}.
The relative bias of factor loadings of the MMLE with rotation deteriorated as $\rho$ increased, which reflects its sensitivity to higher factor correlations, while the BJME was robust to that. For the BJME, intercepts were less biased than factor loadings, which is also consistent with Simulation study 1. 

The average computation time and selected $\lambda$ values over 50 replications are shown in Table~\ref{tab: sim2 time}. The time for the sparse methods corresponds to the elapsed time to select the penalty weight and to estimate the model parameters. 
\begin{table}[htbp]
\caption{Average computation time (seconds) and selected $\lambda$ values over 50 replications in Simulation
study 2}
\label{tab: sim2 time}
\centering
\begin{tabular}{lcccc}
\hline
 & \multicolumn{2}{c}{$N=500$} & \multicolumn{2}{c}{$N=1000$} \\
 \cline{2-5}
 & $\rho=0.1$ & $\rho=0.4$ & $\rho=0.1$ & $\rho=0.4$ \\ 
\hline
BJME & 4.67 & 5.93 & 18.53 & 24.68 \\
 & (14.00) & (14.00) & (14.24) & (16.88) \\
MMLE (sparse inference) & 315.65 & 371.76 & 722.68 & 727.35 \\
 & (0.00076) & (0.00072) & (0.00078) & (0.00052) \\
MMLE (rotation) & 29.61 & 32.86 & 63.75 & 62.31 \\
\hline
\end{tabular}
\end{table}
Table~\ref{tab: sim2 time} indicates that the proposed BJME algorithm is substantially faster than both the sparse MMLE and MMLE with rotation, although the proposed BJME algorithm requires tuning the penalty weight. The sparse MMLE takes longer computation time than the MMLE with rotation because the former also needs to tune the penalty weight. It should be noted that the elapsed time for the sparse MMLE includes computation of the BIC value, which requires the computationally intensive numerical integration, while the proposed CV error metric in Equation~\ref{equ:cv fn} is computationally efficient.
The elapsed time for the BJME was shorter than that of Simulation study 1, although the size of the simulated data is the same. This is because the number of intercepts in Simulation study 1 is larger than that in Simulation study 2, and the dichotomous response setting in Simulation study 2 does not require the strictly ordered constraint for intercepts and reparameterization to accommodate that.

\subsection{Simulation study 3}

\subsubsection{Settings}
% We show that the BJME method has computational efficiency and estimation accuracy even in large-scale and high-dimensional datasets with high latent dimensions. 
In Simulation study 3, the numbers of respondents, items, and latent factors were set to $N=10000$, $J=300$, and $K=\{5, 10\}$. We specified $\rho=.1$, and the other settings were specified in the same manner as in Simulation study 1. As the MCMC method does not scale to such datasets, we applied only the BJME method to the generated datasets.
The generating process of the true values $(\boldsymbol{\Theta}^*,\mathbf{A}^*,\mathbf{D}^*)$ was the same as that used in Simulation study 1. Regarding the starting point, a single start was employed. We used ten cores for parallel computation.

\subsubsection{Results}
\begin{figure}[htp]% [H] is so declass\'e!
\begin{minipage}{0.3cm}
\rotatebox{90}{Rate}
\end{minipage}\hfill
\centering
\begin{minipage}{0.45\textwidth}
\centering
$K=5$\\
\includegraphics[width=\textwidth]{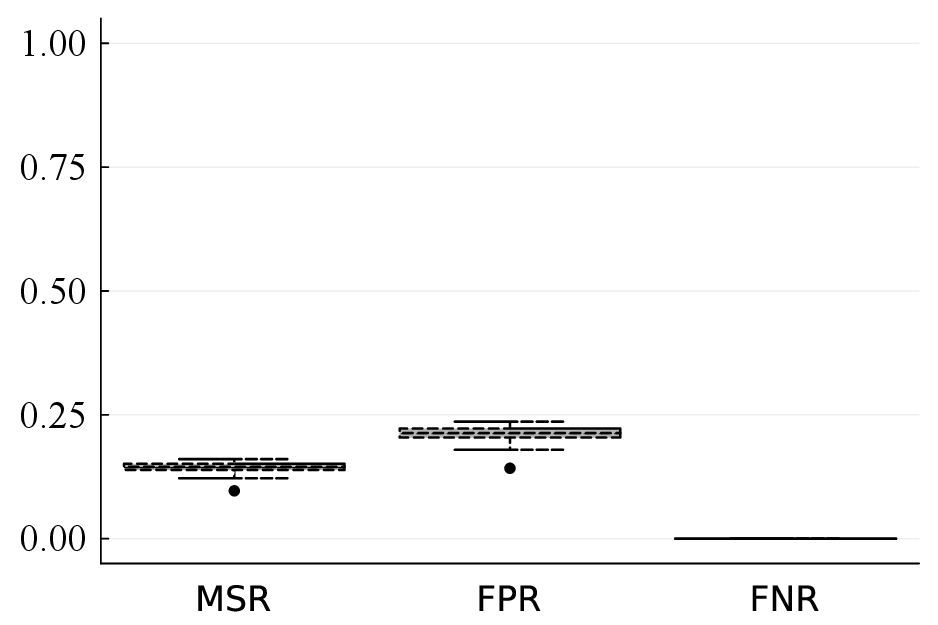}
\end{minipage}\hfill
\begin{minipage}{0.45\textwidth}
\centering
$K=10$\\
\includegraphics[width=\textwidth]{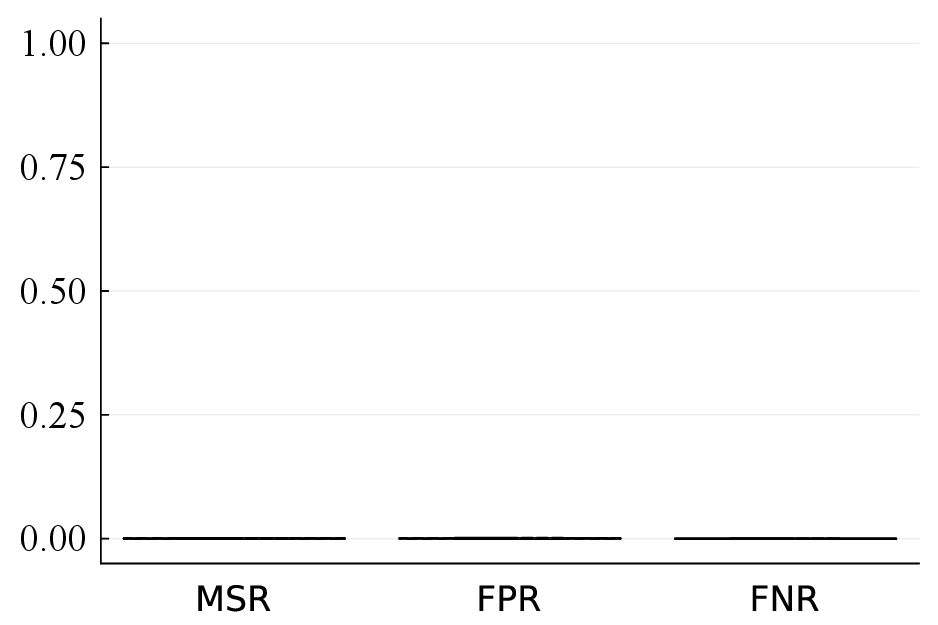}
\end{minipage}
\caption{Selection error in Simulation study 3}\label{fig: sim3 select error}
\end{figure}

Box plots of the error values regarding variable selection over latent factors are presented in Figure~\ref{fig: sim3 select error}.
The MSR, FPR, and FNR values were very small in both $K=5$ and $K=10$. In particular, all these errors were almost zero in $K=10$. The values of FPR were somewhat larger than those of FNR in $K=5$ as well as those in Simulation study 1 and 2. The slightly improved performance in variable selection under $K=10$ was observed because the binary matrix $\btQ^*$ under $K=10$ is more sparser that under $K=5$, meaning that detecting the latent factors with non-zero factor loadings under $K=10$ is relatively easier than the setting of $K=5$.
Note that the difference in the sparsity levels between $K=5$ and $K=10$ is caused by the simulation setting, where the number of latent factors loading on items was at most three under both $K=5$ and $K=10$.
% This difference in the variable selection accuracy between $K=3$ and $K=8$ implies that increasing latent dimensions improves the selection accuracy.

\begin{figure}[htp]% [H] is so declass\'e!
\begin{minipage}{0.3cm}
\rotatebox{90}{Error}
\end{minipage}\hfill
\centering
\begin{minipage}{0.42\textwidth}
\centering
$K=5$\\
\includegraphics[width=\textwidth]{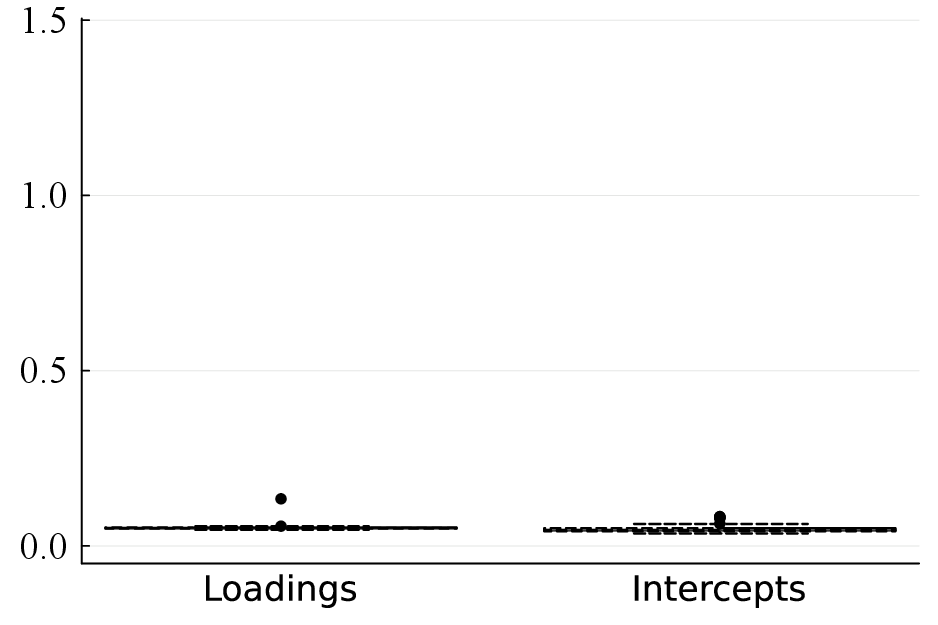}
\end{minipage}\hfill
\begin{minipage}{0.42\textwidth}
\centering
$K=10$\\
\includegraphics[width=\textwidth]{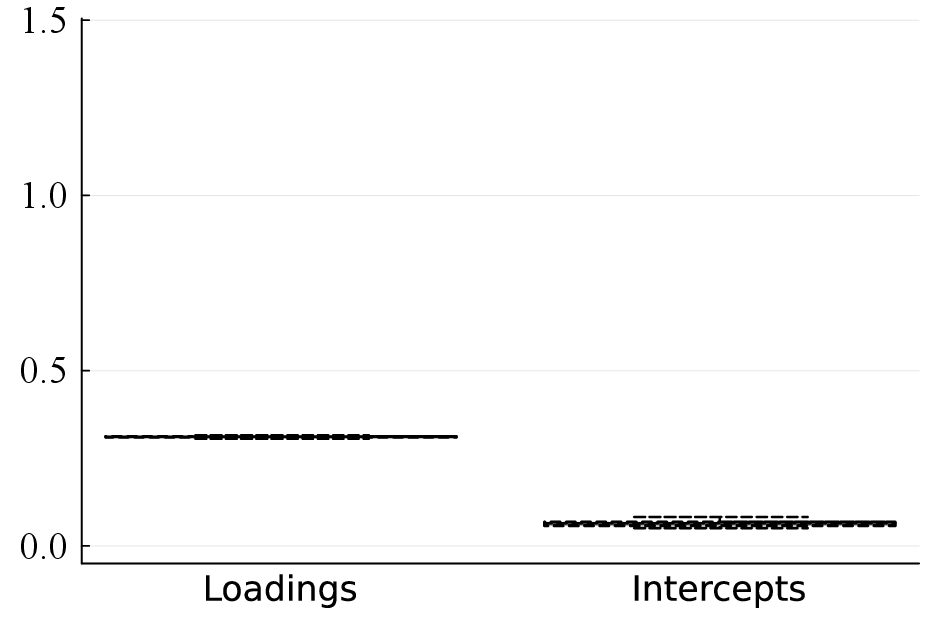}
\end{minipage}\par
\begin{minipage}{0.3cm}
\rotatebox{90}{Relative bias}
\end{minipage}\hfill
\centering
\begin{minipage}{0.42\textwidth}
\centering
$K=5$\\
\includegraphics[width=\textwidth]{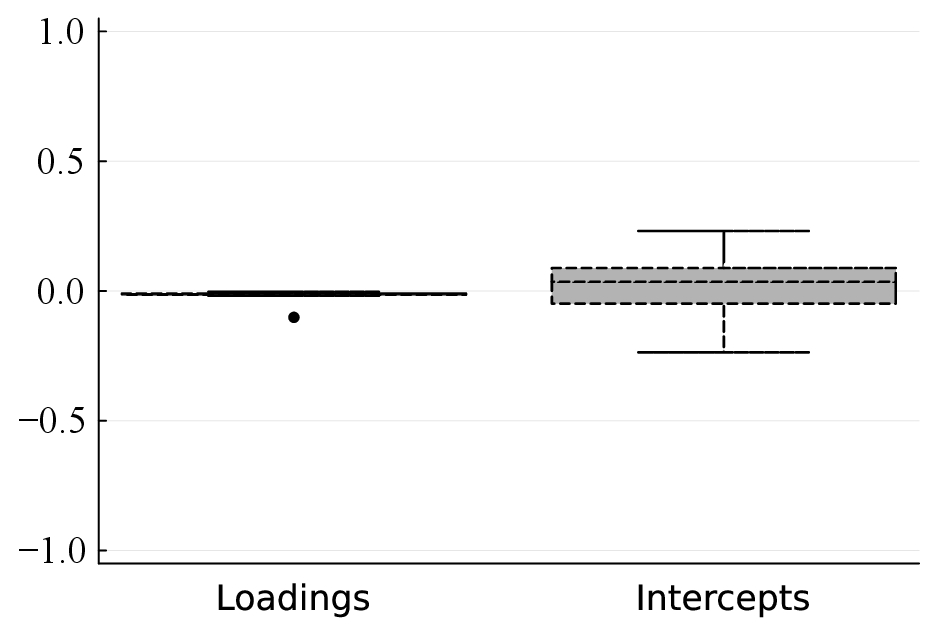}
\end{minipage}\hfill
\begin{minipage}{0.42\textwidth}
\centering
$K=10$\\
\includegraphics[width=\textwidth]{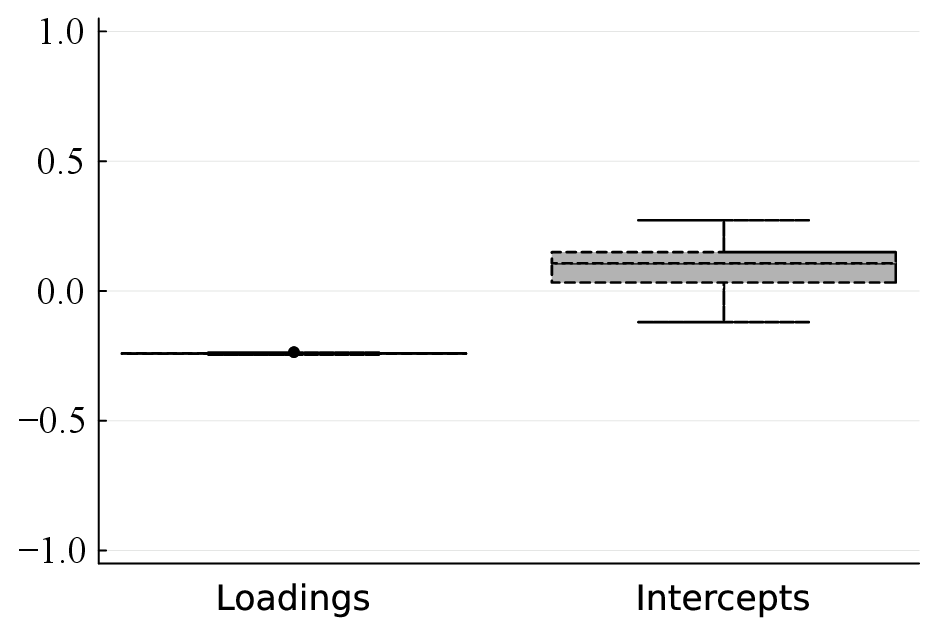}
\end{minipage}
\caption{Recovery error of the model parameters in Simulation study 3}\label{fig: sim3 rmse}
\end{figure}

Next, we investigated the parameter recovery of the model parameters. Box plots of the error regarding the parameter recovery are shown in Figure~\ref{fig: sim3 rmse}.
The estimation of intercepts was slightly more accurate than that of factor loadings, especially in $K=10$, which is consistent with the findings of Simulation study 1 and 2. The absolute values of the errors in the intercept and factor loading parameters were substantially smaller than those in Simulation study 1 and 2. This means that increasing the sample size increases the accuracy of parameter recovery.

The average elapsed time for the CV procedure across replications was 15259.12 and 15353.43 seconds in the conditions of $K=5$ and $K=10$, respectively.
These results show that the BJME method can perform parameter estimation within a reasonable amount of time even in large-scale and high-dimensional settings. Furthermore, when the number of factors was doubled, the computation time increased by much less than a factor of two, which implies that increasing the latent dimensions does not cause a serious computational problem in the BJME method. The average selected value of $\lambda$ was 38.24 and 140.00 under $K=5$ and $K=10$, respectively.

Considering the results of Simulation study 1, the BJME method was slightly less accurate than the MCMC method in terms of the parameter recovery of the model parameters in moderate-scale and low-dimensional settings. However, increasing both the number of respondents and items improved the accuracy of the parameter estimation in the BJME algorithm. Furthermore, increasing the number of respondents and items substantially improved the accuracy of variable selection over latent factors.

\subsection{Simulation study 4}

\subsubsection{Settings}

In Simulation study 4, we varied the number of items as $J = \{50, 100, 150\}$. The number of respondents was set to increase proportionally to the number of items, such that $N = 10J$. We specified $K=3$ and $\rho=0.1$. All other generation processes, the evaluation metrics, and the computational settings were the same as those in Simulation study 3.

\subsubsection{Results}

\begin{figure}[t]
\centering

\begin{minipage}{0.08\linewidth}
\centering
\rotatebox{90}{Rate}
\end{minipage}
\begin{minipage}{0.7\linewidth}
\centering
\includegraphics[width=\linewidth]{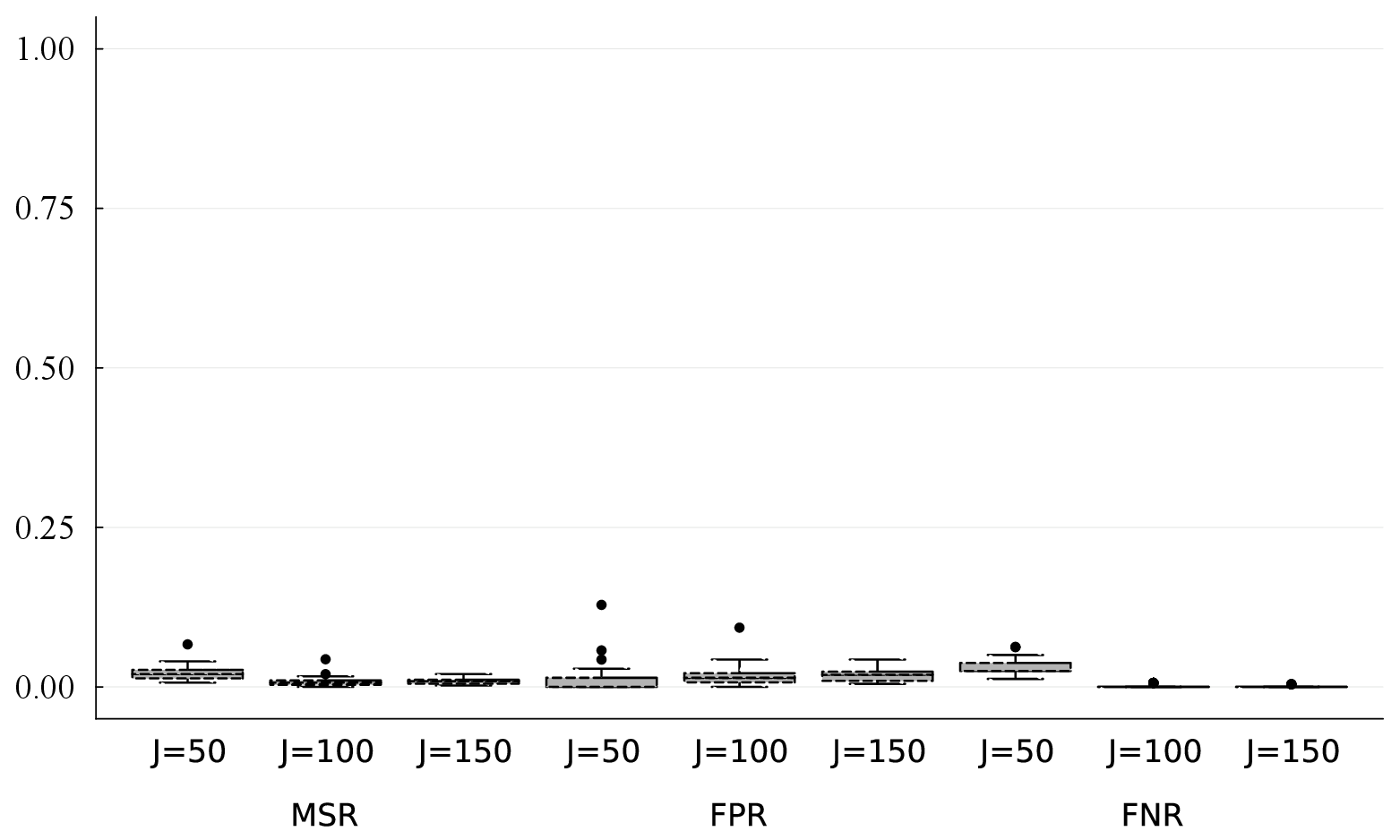}
\end{minipage}
\caption{Selection error in Simulation study 4.}
\label{fig: sim4 select err}
\end{figure}

Figure~\ref{fig: sim4 select err} presents the accuracy of variable selection over latent factors in terms of MSR, FPR, and FNR. The results indicated that the proposed BJME algorithm successfully recovered the sparse loading structure across all conditions. As the number of respondents and items increased, the MSR decreased. Both the FPR and the FNR remained low across all conditions, which suggests that the proposed algorithm effectively controls both types of selection errors when the dimensionality of the problem increases.

\begin{figure}[t]
\centering

\begin{minipage}{0.08\linewidth}
\centering
\rotatebox{90}{Error}
\end{minipage}
\begin{minipage}{0.7\linewidth}
\centering
\includegraphics[width=\linewidth]{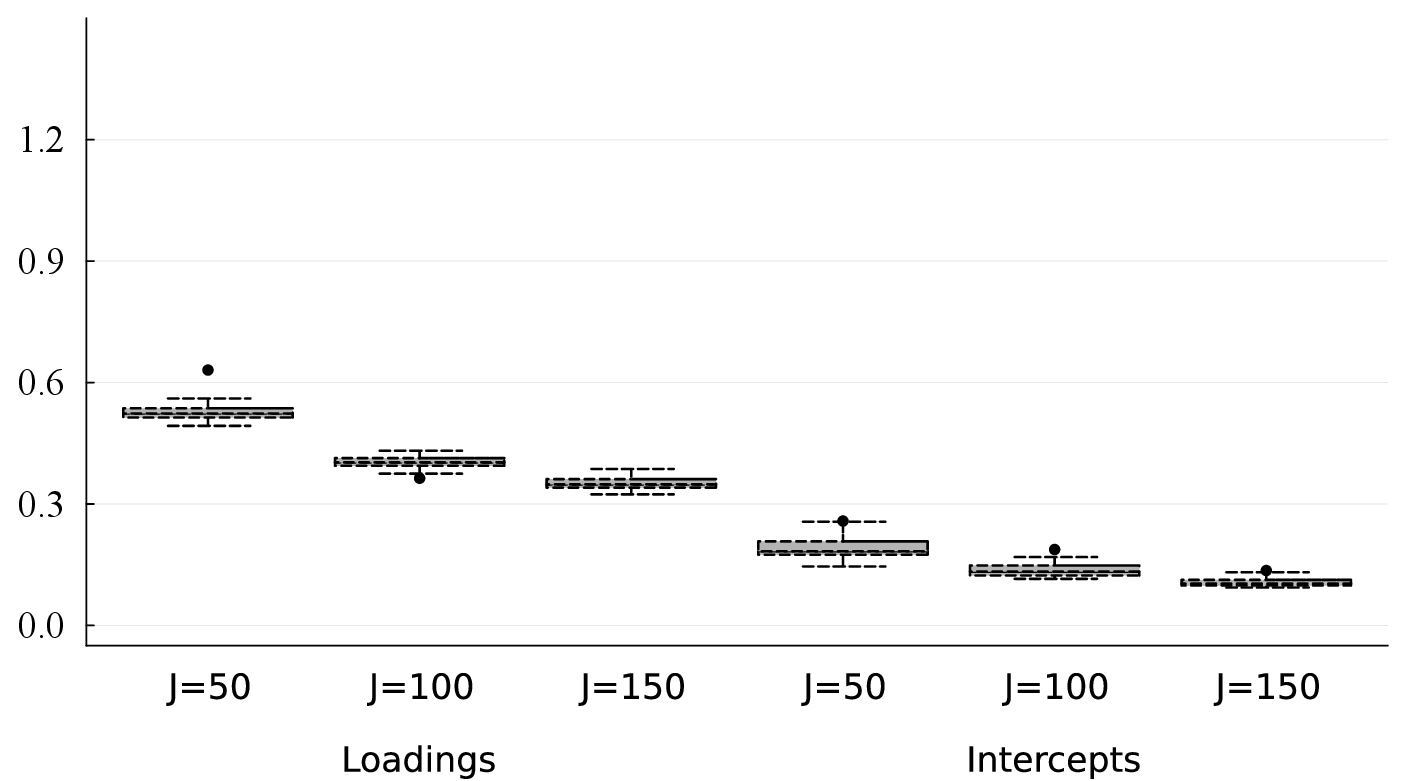}
\end{minipage}

\vspace{0.2cm}

\begin{minipage}{0.08\linewidth}
\centering
\rotatebox{90}{Relative bias}
\end{minipage}
\begin{minipage}{0.7\linewidth}
\centering
\includegraphics[width=\linewidth]{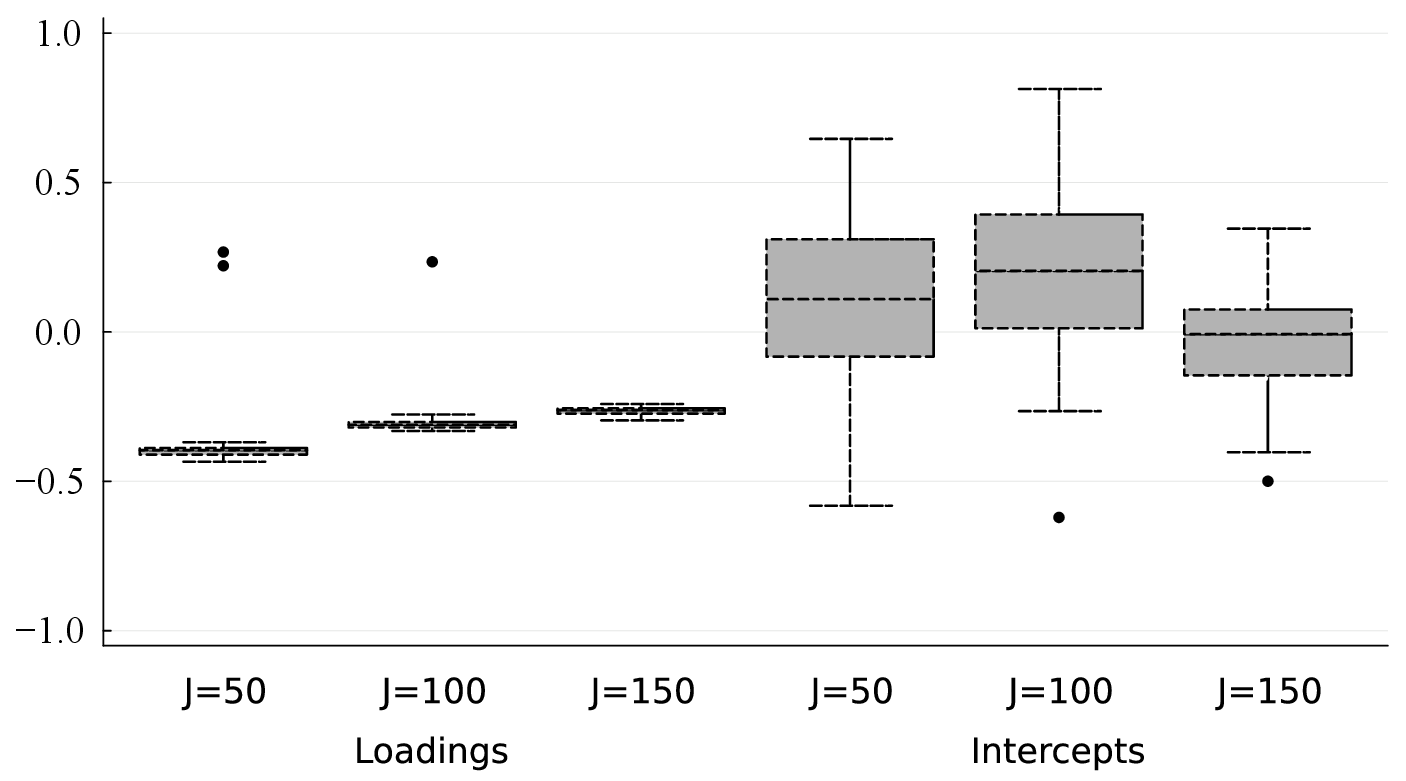}
\end{minipage}

\caption{Recovery error of the model parameters in Simulation study 4}
\label{fig: sim4 est err}
\end{figure}

Figure~\ref{fig: sim4 est err} summarizes the estimation accuracy of the model parameters. As both $N$ and $J$ increased, the estimation errors of factor loadings and intercepts decreased. In particular, the error of loadings decreased and the relative bias became smaller as the dataset size increased. These results suggest that the proposed algorithm benefits from larger datasets. For the intercept parameters, the error also decreased as $N$ and $J$ increased. However, the relative bias of the intercept parameters remained largely unchanged across the conditions. Comparing the estimation accuracy of loadings and intercepts, intercepts were estimated more accurately than loadings, which is the same trend as the previous simulation studies.

Overall, these results demonstrate that the proposed BJME algorithm maintains accurate recovery of the sparse loading structure and improves parameter estimation accuracy as the dataset size increases. 
This finding is consistent with previous findings on JMLE methods \parencite{chen2019joint,chen2020structured}, which established statistical consistency of item parameter estimators when the numbers of respondents and items go to infinity simultaneously.
% It is consistent with existing parameter inference methods based on the JMLE \parencite{chen2019joint,chen2020structured} that parameter recovery improves when the number of respondents and items increases simultaneously.

We also examined the computational cost of the proposed algorithm as the number of respondents and items increased. The average computation time for the CV procedure was 10.62, 70.62, and 279.88 seconds for $J=50, 100, 150$, respectively. These results indicate that the computation time increases as the size of an item response matrix grows, but remains manageable for the considered settings. The average selected value of $\lambda$ was 13.52, 13.76, and 14.00, respectively for $J=50, 100, 150$.

\section{Real data analysis} \label{sec:5}
We illustrate the utility of the proposed algorithm in investigating the relationships between items and latent factors using an empirical dataset.
Specifically, we analyzed the synthetic aperture personality assessment (SAPA) personality inventory (SPI) dataset \parencite{condon2018sapa,REVELLE2021exploring}, which can be loaded using the \texttt{psychTools} package \parencite{revelle2023psych} in R. The SPI dataset comprises 4,000 respondents, 135 items, and 6 response categories (i.e., “Very Inaccurate,” “Moderately Inaccurate,” “Slightly Inaccurate,” “Slightly Accurate,” “Moderately Accurate,” and “Very Accurate”) and has no missing entries. The items were primarily selected from the international personality item pool, and 70 items out of 135 were designed to measure the Big Five factors; we used these 70 items in the real data analysis. When analyzing the dataset, we rearranged the order of the items to better interpret the relationships between the items and latent factors. In particular, the items were rearranged so that 1--14, 15--28, 29--42, 43--56, and 57--70 were related to agreeableness, conscientiousness, neuroticism, extraversion, and openness, respectively. We then reverse-scored the responses for the reversed items.

Furthermore, we considered the number of latent factors to be known and set it to $K=5$. We tuned the penalty weight from a set of candidate values and plugged in the covariance matrix of the latent factors in the same manner as in the simulation studies. We used a single starting point, and the starting point was generated in the same manner as in the simulation studies.
Regarding computation, we used ten cores on the same platform as that used in the simulation studies. The elapsed time for the CV procedure was 1624.79 seconds, and the selected value of $\lambda$ was 38.0.

\begin{figure}[htp]% [H] is so declass\'e!
\begin{minipage}{0.3cm}
\rotatebox{90}{Item}
\end{minipage}\hfill
\centering
\begin{minipage}{0.42\textwidth}
\centering
$\mathbf{Q}^*$\par\medskip
\includegraphics[width=\textwidth]{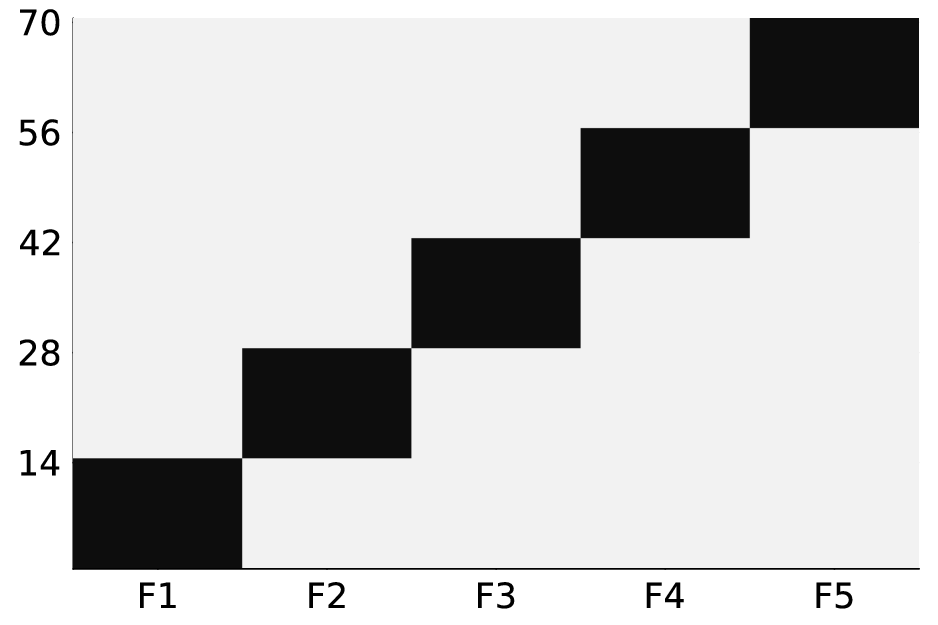}
\end{minipage}\hfill
\begin{minipage}{0.42\textwidth}
\centering
$\hat{\mathbf{A}}$ in absolute value\par\medskip
\includegraphics[width=\textwidth]{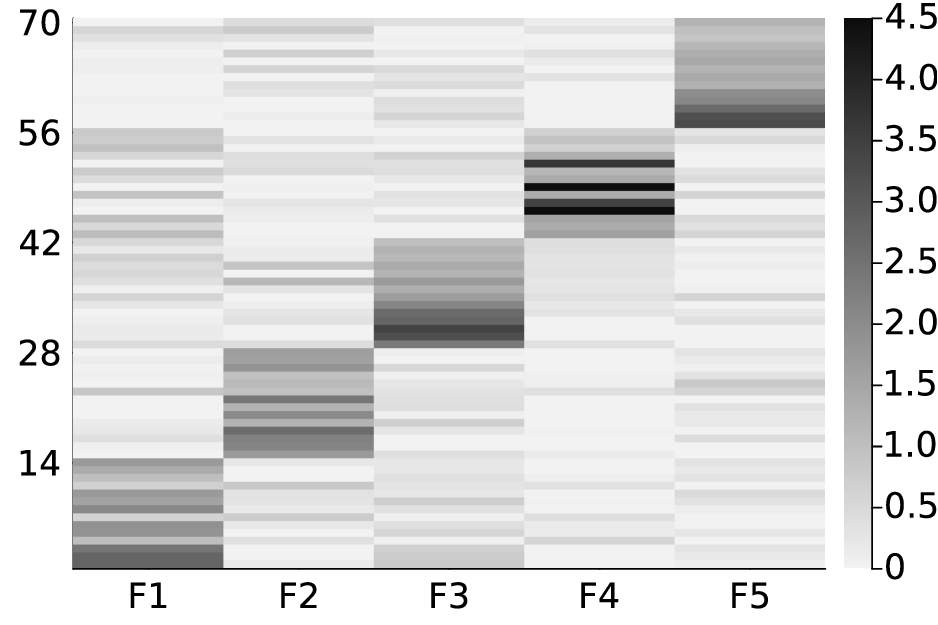}
\end{minipage}
\caption{Heatmaps of the sparse structure determined by the assessment design and estimated factor loadings in absolute value}\label{fig:heatmap of factor loadings}
\end{figure}

To assess whether the proposed algorithm successfully uncovers the sparse structure consistent with the assessment design, we compared the estimated factor loadings $\hat{\mathbf{A}}$ with the sparse structure $\mathbf{Q}^*$ determined by the assessment design of SPI. The heatmaps of $\mathbf{Q}^*$ and the estimated factor loading structure are presented in Figure~\ref{fig:heatmap of factor loadings}.

The white and black blocks on the left-hand side of Figure~\ref{fig:heatmap of factor loadings} correspond to the entries of $q_{jk}^*=0$ and $q_{jk}^*=1$, respectively. The five latent factors are labeled F1, F2, F3, F4, and F5, which correspond to agreeableness, conscientiousness, neuroticism, extraversion, and openness, respectively. Note that we addressed column permutation and column sign flips using the designed $\mathbf{Q}^*$ after obtaining the estimates.
Compared to the sparse structure presented on the left-hand side, the absolute values of the estimated factor loadings take higher values in the entries where $q_{jk}^*$ equals one, and vice versa.
In fact, the point-biserial correlation coefficient between the vectorized absolute values of $\hat{\mathbf{A}}$ and vec$(\mathbf{Q}^*)$ was 0.83. These results suggest that the proposed algorithm provides a factor loading structure that is interpretable and consistent with the assessment design.

Note that not all estimated factor loadings take the value of zero in the entries with $q_{jk}^*=0$. However, some psychological studies have suggested that empirical datasets do not have perfect simple structures such as $\mathbf{Q}^*$ in Figure~\ref{fig:heatmap of factor loadings} but can have many cross-loadings \parencite{marsh2013measurement,booth2014exploratory}. Moreover, to evaluate the magnitude of the estimated factor loadings in entries with $q_{jk}^*=0$, we computed the mean of the values of the estimated factor loadings in entries with $q_{jk}^*=0$:
\begin{align}
\frac{\sum_{j=1}^J\sum_{k=1}^K\mathbb{1}\{q_{jk}^*=0\} |\hat{a}_{jk}| }{\sum_{j=1}^J\sum_{k=1}^K\mathbb{1}\{q_{jk}^*=0\}}=0.22,
\end{align}
which is sufficiently small to ignore compared to the mean of the values of the estimated factor loadings in entries with $q_{jk}^*=1$:
\begin{align}
\frac{\sum_{j=1}^J\sum_{k=1}^K\mathbb{1}\{q_{jk}^*=1\} |\hat{a}_{jk}| }{\sum_{j=1}^J\sum_{k=1}^K\mathbb{1}\{q_{jk}^*=1\}}=1.82.
\end{align}
These quantities suggest that the proposed algorithm shrinks the factor loadings of irrelevant latent factors to negligibly small values.

% ヒートマップをみると、真のＱ行列の構造を抽出できていることが分かる。相関も～と高い値であった。
% 真のQ行列が0の箇所は、0に縮小できなくとも、0.5とかは超えない値で推定されている。この程度の値は、反応確率を～程度にしか上昇させないため、無視できるものと考えられる。

\section{Discussion} \label{sec:6}
This study revisited BJME, a classical Bayesian estimation approach, and shed new light on its utility in performing regularized estimation in exploratory IFA and its potential as a computationally efficient alternative to the MCMC method. In particular, using the correspondence between prior distributions and penalty functions, we developed a sparse BJME algorithm under the MGRM that performs the sparse estimation of factor loadings and has high computational efficiency owing to the gradient-based optimization method that is easy to implement and compute. Simulation studies showed that the proposed algorithm has superior computational efficiency over the MCMC and MMLE methods and scales well, even to large-scale and high-dimensional datasets. Moreover, the proposed algorithm is easy to parallelize, which substantially improves the computational efficiency. Simulation studies also showed that the proposed algorithm has high accuracy in parameter recovery of the model parameters and variable selection over latent factors, especially in large-scale and high-dimensional settings. In addition, we illustrated the use of the proposed algorithm on the Big Five personality questionnaire dataset. The results in the real data analysis revealed that the proposed algorithm can extract an interpretable factor loading structure with high computational efficiency in parameter estimation.
%\suggest{You probably need to add what you achieved in the real data analysis a little bit more to emphasize the contribution of this study.}}

Although this study focused on exploratory IFA under the MGRM, the proposed algorithm can be easily modified for a confirmatory case of the MGRM. In the confirmatory case, the true sparse structure $\mathbf{Q}^*$ is known prior to the analysis. Therefore, we can perform confirmatory analysis by constraining $a_{jk}=0$ where $q_{jk}^*=0$ and changing the prior for factor loadings to normal distributions.
Furthermore, in this study, we plugged in the covariance matrix of the latent factors $\boldsymbol{\Sigma}_\theta$ instead of estimating it. However, we can incorporate the estimation of $\boldsymbol{\Sigma}_\theta$ into the current algorithm. 
Estimating a correlation matrix may be computationally unstable because it requires that the matrix be positive definite and its diagonal entries all be one. 
%\suggest{you need positive definite. semi-definite is not enough because semi-definite allows zer eigenvalues, meaning that the determinant can be zero. Thus, it is not invertible.}. 
To achieve these constraints, we can employ the reparameterization technique of the Cholesky decomposition of $\boldsymbol{\Sigma}_\theta$ for computational stability in its optimization.

% 今回はprediction-optimalなCVだったので、Q行列の正解率は良くなかった。これは、小さめのpenalty weightを選んでしまったため(CVのやり方の問題)。model selection optimalを狙うのであればCVでdebiasingも行うべき(murphy)。

% しかしながらそもそも推定量の性質が不透明なので、そこも考える必要がある。

In the comparison of the estimation accuracy among the BJME, MCMC, and MMLE methods, the results of Simulation study 1 and 2 showed that the BJME method is slightly less accurate than the MCMC or MMLE methods in middle-scale and low-dimensional settings. Considering that BJME is a Bayesian counterpart of JMLE, this result is consistent with the findings of \textcite{chen2019joint}, which found that the MMLE provides more accurate estimates than JMLE in moderate-scale and moderate-dimensional datasets such as $N=\{1000,2500\}$ and $J=100$. 
This is because the likelihood function of the latent variable models in the MCMC methods is identical to that in the MMLE method, which targets the marginal likelihood function. 
%Similar to EM algorithms in the MMLE, the marginalization of latent variables in the likelihood function is addressed by the DA algorithm \parencite{brooks2011handbook}, and the posterior distribution of latent variables is obtained as a byproduct of the DA algorithm. 
We plan to further investigate the relationship between the BJME and MCMC methods in future research.

Regarding the selection of the penalty weight $\lambda$, our CV method obtained a prediction-optimal value of $\lambda$ in the sense that we chose the value of $\lambda$ minimizing the prediction error in Equation~\ref{equ:cv fn}. This type of CV tends to pick a value of $\lambda$ that is not large enough to make irrelevant factor loadings actually zero. 
Consequently, a less sparse model is selected \parencite[][Section 13.3.5]{murphy2012machine}. Although simulation studies show that the proposed algorithm provides sufficient performance in finding the true sparse structure, a prediction-optimal value of $\lambda$ is not necessarily model-selection optimal. In the latter case, \textcite{murphy2012machine} noted that debiasing the Lasso estimates is preferable to obtain better estimates of non-zero factor loadings because the corresponding Lasso estimates tend to shrink toward zero compared with the corresponding true values. Similarly, extensions to the adaptive Lasso \parencite{zou2006adaptive} or hierarchical adaptive Lasso \parencite{lee_hierarchical_2010} are promising directions for future research. Furthermore, we regarded the number of latent factors as known in this study. However, a method for selecting its value can still be considered. Our CV method would be employed to select both the number of factors and the penalty weight using a grid search, as in \textcite{lee2010sparse}.

In conclusion, the sparse BJME algorithm is highly computationally efficient and achieves high accuracy in variable selection over latent factors and in recovering the model parameters in both large-scale and high-dimensional datasets. 
Therefore, the BJME method is a promising Bayesian estimation approach in this era of big data.

% Because the theoretical properties of the estimators by the proposed method are unclear.
% あと次元数選択の話、Chen 2021 psychometrika

\section*{Acknowledgements}
The authors used ChatGPT to assist with language editing and with the development of code, mainly for generating figures. All statistical analyses, results, and visualizations were conducted and verified by the authors. This work was supported by JST PRESTO Grant Number JPMJPR21C3 and JSPS KAKENHI Grant Numbers 25H00568, 25K06746.

\printbibliography

\section*{Appendix}
In this section, we provide detailed derivations and explanations for the update rules presented in Equations~\ref{equ:theta update}, \ref{equ:a update}, and \ref{equ:d update}. These updates form the core of the proposed BJME algorithm. Below, to avoid notational clutter, we use the following shorthand:
\[
P_{ij}(c) = P(Y_{ij} \ge c|\boldsymbol{\theta}_i, \mathbf{a}_j, \mathbf{d}_j).
\]
This represents the conditional probability that respondent $i$ answers item $j$ with category $c$ or higher.

The gradient for $\boldsymbol{\theta}_i$ is given by:
\begin{align}
\frac{\partial \ell(\bTheta, \btA, \btD\mid \btY, \lambda)}{\partial\boldsymbol{\theta}_{i}} = &\sum_{j=1}^{J} \sum_{c=0}^{C_{j}-1} \mathbb{1}\{Y_{ij}=c\} \frac{P_{ij}(c)(1 - P_{ij}(c)) - P_{ij}(c+1)(1 - P_{ij}(c+1))}{P_{ij}(c) - P_{ij}(c+1)} \mathbf{a}_j \\
&- \mathbf{\Sigma}_{\theta}^{-1}\boldsymbol{\theta}_{i}.
\end{align}

Regarding category intercepts $\mathbf{d}_j$, the gradient of $d_{jc}$ is given by
\begin{align}
\frac{\partial \ell(\boldsymbol{\Theta}, \mathbf{A}, \mathbf{D} \mid \mathbf{Y}, \lambda)}{\partial d_{jc}} &= \sum_{i=1}^N \mathbb{1}\{Y_{ij}=c\}\frac{P_{ij}(c)(1-P_{ij}(c))}{P_{ij}(c)-P_{ij}(c+1)} \\
& \quad - \sum_{i=1}^N \mathbb{1}\{Y_{ij}=c-1\}\frac{P_{ij}(c)(1-P_{ij}(c))}{P_{ij}(c-1)-P_{ij}(c)} - \frac{d_{jc}}{\sigma_d^2}.
\end{align}
According to the reparameterization rule, the gradient of $\boldsymbol{\delta}_{j}$ is derived based on the gradient of $d_{jc}$ as
\begin{align}
\frac{\partial \ell(\boldsymbol{\Theta}, \mathbf{A}, \mathbf{D} \mid \mathbf{Y}, \lambda)}{\partial \delta_{j1}} &= \sum_{c=1}^{C_j-1} \frac{\partial \ell(\boldsymbol{\Theta}, \mathbf{A}, \mathbf{D} \mid \mathbf{Y}, \lambda)}{\partial d_{jc}}, \\
\frac{\partial \ell(\boldsymbol{\Theta}, \mathbf{A}, \mathbf{D} \mid \mathbf{Y}, \lambda)}{\partial \delta_{jc^\prime}} &= -\exp(\delta_{jc^\prime}) \sum_{c=c^\prime}^{C_j-1} \frac{\partial \ell(\boldsymbol{\Theta}, \mathbf{A}, \mathbf{D} \mid \mathbf{Y}, \lambda)}{\partial d_{jc}},
\end{align}
for $c^\prime=2,\dots,C_j-1$.

Furthermore, the gradient with respect to $\mathbf{a}_j$ is written as
\begin{align}
\frac{\partial\log P( \btY \mid \bTheta, \btA, \btD)}{\partial \textbf{a}_j} = \sum_{i=1}^{N} \sum_{c=0}^{C_{j}-1} \mathbb{1}\{Y_{ij}=c\} \frac{P_{ij}(c)(1 - P_{ij}(c)) - P_{ij}(c+1)(1 - P_{ij}(c+1))}{P_{ij}(c) - P_{ij}(c+1)} \boldsymbol{\theta}_i.
\end{align}
These derivations provide the detailed steps needed to implement and follow the sparse BJME algorithm accurately.

% \section*{Authors}
% KEIICHIRO HIJIKATA is a former master's student in the Graduate School of Education at the University of Tokyo, Tokyo, Japan; email: k.hijikata.1120@outlook.jp. His research interests are in psychometrics.

% MOTONORI OKA is a doctoral student in the Department of Statistics at the London School of Economics and Political Science, London, UK; email: m.oka1@lse.ac.uk. His research interests are in psychometrics.

% KENSUKE OKADA is an associate professor in the Graduate School of Education at the University of Tokyo, Tokyo, Japan; email: ken@p.u-tokyo.ac.jp. His research interests are in psychometrics and Bayesian statistics.

\end{document}